\documentclass[preprint]{JHEP3} 

\JHEPspecialurl{http://jhep.sissa.it/JOURNAL/JHEP3.tar.gz}

\usepackage{epsfig,multicol}

\def\beq{\begin{equation}}

\def\eeq{\end{equation}}
\def\eq#1{{Eq.~(\ref{#1})}}

\newcommand{\bea}{\begin{eqnarray}}

\newcommand{\eea}{\end{eqnarray}}

\parskip=\itemsep               

\newcommand{\beqar}[1]{\begin{eqnarray}\label{#1}}

\newcommand{\eeqar}{\end{eqnarray}}

\def\thefootnote{\fnsymbol{footnote}} 

\title{
{\Large  \bf One gluon, two gluon: multigluon production via high energy evolution}}
\author{A. Kovner and M. Lublinsky   \\
Physics Department, University of Connecticut, \\ 2152 Hillside
Road, Storrs, CT 06269-3046, USA}

\abstract{We develop an approach for calculating the inclusive multigluon production within the JIMWLK high energy evolution. We give a formal expression of multigluon cross section in terms of a generating functional for arbitrary number of gluons $n$. In the dipole limit the expression simplifies dramatically. We recover the previously known results for single and double gluon inclusive cross section and generalize those for arbitrary multigluon amplitude in terms of Feynman diagramms  of Pomeron - like objects coupled to external rapidity dependent field $s(\eta)$. We confirm the conclusion that the AGK cutting rules in general are violated in multigluon production. However we present an argument to the effect that for doubly inclusive cross section the AGK diagramms give the leading contribution at high energy, while genuine violation only occurs for triple and higher inclusive  gluon production. We discuss some general properties of our expressions and suggest a line of argument to simplify the approach further.

}

\begin{document}



\def\thefootnote{\arabic{footnote}} 

\section{Introduction}

The study of high energy hadronic scattering and the physics of saturation has been a very active area of research in recent years\cite{pomeron, dipole, mv, balitsky, Kovchegov, JIMWLK, cgc}. The effort has  bifurcated recently with one branch concentrating on understanding dynamics of Pomeron loops \cite{pomloops, kl, something, kl1, kl4, kl5} while the other one pushing towards better understanding of more detailed aspects of JIMWLK evolution\cite{JIMWLK,cgc}, such as relation to the QCD Regge theory \cite{7P, YY} and calculation of observables less inclusive than the total cross section\cite{JMK, BKNW, HWS, klw}. The latter line of research seems especially relevant since we understand by now that the total cross section cannot be calculated perturbatively at asymptotically high energies, even if the Pomeron loops are taken into account \cite{froissart}.

The present paper is devoted to a detailed discussion of multigluon production in the JIMWLK approach. 
The study of gluon production at high energy was initiated in a paper by Kovchegov and Tuchin \cite{Kovchegov:2001sc} which calculated the single gluon inclusive cross section within the dipole model approximation\cite{dipole}. This was later extended beyond the dipole model \cite{Blaizot:2004wu},\cite{BKNW}. The expression for the double gluon inclusive production in the dipole limit  with an arbitrary rapidity interval between the two gluons
was obtained by a real {\it tour de force} calculation in \cite{JMK}, while a simpler calculation of two gluon production at the same rapidity but beyond the large $N_c$ limit was performed in \cite{BKNW}. 
The double gluon inclusive cross section was also calculated by Braun \cite{braun} in the framework of the BFKL Pomeron \cite{BFKL}
calculus. One should note that the results of \cite{JMK} and \cite{braun} are different, as the AGK cutting rules \cite{AGK} 
are violated in \cite{JMK} while \cite{braun} suggests that the AGK violating contributions are small at high energy. 

All the above calculations have been performed on a case by case basis, and at the moment there is no unified approach that would streamline calculations of an arbitrary observable of this type. Recently \cite{klw} we have introduced such a unified formalism for calculation of semi-inclusive observables such as various diffractive and elastic cross sections. In the present paper our goal is to extend \cite{klw} to calculation of multigluon production.

The structure of this paper is the following. 
In Section 2 we recall the general formalism of the high energy evolution as it arises from the evolution of hadronic wave function and applies to semi-inclusive observables. In Section 3 we extend the formalism to include multigluon production. We write down a formal expression for the generating functional which generates the inclusive probability of production of $n$ gluons at arbitrary rapidities $Y_i$ and arbitrary transverse momenta $k_i$. We discuss some properties of this expression and show that it reproduces the known result for the single gluon inclusive case.

In Section 4 we specify to the dipole model limit in which the problem simplifies considerably. In this limit we can develop a simple diagrammatic technique in terms of the propagators of BFKL Pomeron-like objects propagating in the external field of the dipole $s$-matrix $s(\eta)$. We confirm that for double inclusive gluon production our result agrees with \cite{JMK}. We stress that not only the BFKL Pomeron but also other similar objects propagate in the relevant diagrams if the number of produced gluons is greater than one. It is the presence of these additional objects that violates the AGK cutting rules. 
The actual calculation of the multigluon cross section requires the knowledge of the propagators of these "cousins" of the BFKL Pomeron. We find the asymptotic behavior of these propagators for small and large separations and show that the crossover must occur on the scale of the inverse saturation momentum of the target. Based on the behavior of the propagator and of the gluon emission vertices we argue that the leading high energy contribution to double inclusive gluon production still preserves the AGK cutting rules thus reconciling the results of \cite{JMK} and \cite{braun}. We show however that for triple and higher gluon production the AGK rules are genuinely violated. We formulate an ansatz which incorporates the known asymptotics of the relevant propagators with the correct crossover scale which we believe should be a good approximation to physics. We hope that numerical calculations using this ansatz may be feasible.

\section{High energy scattering: the general setup}

\subsection{Evolution of a wavefunction}
In this section we recall the high energy evolution of the hadronic wave function as discussed in 
\cite{kl, kl1, something, YY}.

Consider an arbitrary projectile whose wave function in the gluon Fock space can be written as
\begin{equation}
|\Psi\rangle_{Y_0}\,=\,\Psi[a^{\dagger a}_i(x)]|0\rangle_{Y_0}\,.
\label{wf}
\end{equation}
The gluon creation operator $a^\dagger$ depends on the transverse coordinate, and also on the longitudinal momentum $k^+$. 
The projectile is assumed to have large energy. The gluon operators which enter eq.(\ref{wf}) therefore all have longitudinal
 momenta above some cutoff $\Lambda$. We will refer to these degrees of freedom as "valence", $|\Psi\rangle=|v\rangle$.
Henceforth we omit the dependence on longitudinal momentum in our expressions, as the momentum enters only as a spectator
 variable and only determines the total phase space available for the evolution.
The wave function in eq.(\ref{wf}) is arbitrary. 

When boosted, the valence wave function gets dressed by an evolution operator $\Omega$ \cite{kl, zakopane}, which depends
on the valence degrees of freedom ($\rho$) as well as on newly created soft modes ($a$).
 The evolved wave function has the following structure 
\beq
|\Psi \rangle_Y\,=\, \Omega_Y(\rho,\,a)\,|v\rangle_{Y_0}\,; \ \ \ \ \ \ \ \ \ \ \  \ \ \ \ \ \ 
|v\rangle_{Y_0}\,=\,|v\rangle\,\otimes\,|0_a\rangle
\eeq
where $|0_a\rangle$ is the Fock vacuum of the Hilbert space of the soft modes.
The evolution operator $\Omega$ is explicitly known in the dilute limit ($\rho\rightarrow 0$) only. In this limit,
$\Omega\,=\,C$, where $C$ is  the coherent operator 
\beq\label{co}
C_Y\,=\,{\rm Exp}\left\{
i\,\int d^2x\,b_i^a(x)\int_{e^{Y_0}\,\Lambda}^{e^Y\,\Lambda}{dk^+\over \pi^{1/2}| k^+|^{1/2}} 
\left[a^{ a}_i(k^+, x)\,+\,a^{\dagger a}_i(k^+, x))\right]\right\}\,.
\eeq
The creation operators $a^\dagger(k^+)$ create gluons with soft momenta, which are not present in the valence
 state $|v\rangle$. The field $b$ depends only on the valence degrees of freedom. It is determined as the solution of
 the "classical" equation of motion 
\begin{eqnarray}
&&\partial_ib_i^a\,+\,g\,\epsilon^{abc}\,b^b_i(x)\,b^c_i(x)\,=\,g\,\rho^a(x)\nonumber\\
&&\epsilon_{ij}\,[\partial_ib^a_j\,-\,\partial_jb^a_i\,+\,g\epsilon^{abc}b^b_i\,b^c_j]\,=\,0
\label{b}
\end{eqnarray}
where $\rho^a(x)$ is the color charge density operator of the projectile. The coherent operator $C$
dresses the valence wave function by a cloud of the Weiszacker-Williams gluons. Operatorially
\begin{equation}\label{shift1}
C^{\dagger}\, A^a_i(k^+, x)\,C\,=\,A^a_i(k^+, x)+{i\over k^+}b_i^a(x)\,.
\end{equation}

The evolution of a generic non-diagonal matrix element of any observable $\hat{\cal O}$ is given by
\beq
\frac{d\,\langle v|\hat{\cal O}|v'\rangle}{d\,Y}\,=\,\lim_{{Y\rightarrow Y_0}}
\frac{\langle v|\Omega^\dagger_Y\, \hat{\cal O}(\rho+\delta\rho)\,\Omega_Y|v'\rangle\,\,
-\,\,\langle v|\hat{\cal O}(\rho)|v'\rangle}{Y\,-\,Y_0}\,=-\,\langle v|\,H[\rho]\,\,\,\hat{\cal O}(\rho)\,|v'\rangle\,.
\eeq
Here both states $|v\rangle$ and $|v'\rangle$ are boosted by the same amount.
With the explicit form of the evolution operator $C$ the evolution Hamiltonian is obtained as 
\beq
H^{KLWMIJ+}[\rho,{\delta\over\delta\rho}]\,=\,
\int{d^2z \over 2\,\pi}\, \tilde Q^a_i(z)\,\tilde Q^a_i(z)\,
\end{equation}
where the one gluon emission amplitudes $Q^a_i(z)$ is defined as 
\begin{equation}\label{Q}
\tilde Q^a_i(z)\,=\,R^{ab}(z)\,b^b_{R\,i}(z)\,-\,b^a_{L\,i}(z)\,,
\label{qus}
\end{equation}
with the "dual Wilson line" $R$ 
\begin{equation}
R(z)^{ab}\,\,=\,\,
\left[{\cal P}\exp{\int_{0}^{1} d z^-\,T^c\,{\delta\over\delta\rho^c(z,\,z^-)}}\right]^{ab}\,.
\label{rr}
\end{equation}
Here $b_{L,R}$ are solutions of the classical equations of motion with sources given  by
the generators of the left/right color rotations $\bar J_{L,R}$ defined as
\begin{equation}
\bar J_R^a(x)=-{\rm tr} \left\{R(x)T^{a}{\delta\over \delta R^\dagger(x)}\right\}, \ \ \ \  \bar J_L^a(x)=
-{\rm tr} \left\{T^{a}R(x){\delta\over \delta R^\dagger(x)}\right\}, \ \ \ \  \ \ \ \ \ \ \ 
\bar J_L^a(x)\,\,=\,\,[R(x)\, \bar J_R(x)]^a\,.
\end{equation}
The rotations have an  alternative representation in terms of the charge density operators:
\begin{equation}
\bar J_R^a(x)\,=\,\rho^a(x,x^-=1); \ \ \ \ \ \ \ \ \ \ \ \ \ \ \ \bar J_L^a(x)\,=\,\rho^a(x,x^-=0)\,.
\label{jays}
\end{equation}
The KLWMIJ+ Hamiltonian  reduces to the KLWMIJ Hamiltonian when we keep the linear approximation for the classical field $b$: 
\begin{equation}
b^a_i(z)\,=\,{g\over 2\pi}\int d^2x{(z-x)_i\over (z-x)^2}\,\rho^a(x)\,.
\end{equation} 
and then 
\begin{eqnarray}\label{KL}
\tilde Q^a_i(z)\,=\,\int\, d^2x{(x-z)_i\over (x-z)^2}\,[R^{ab}(z)\,-\,R^{ab}(x)]\,\bar J^b_R(x)\,.
\end{eqnarray}
It is the last form that we will use explicitly throughout this paper although generalization to KLWMIJ+ does not present any difficulties.

Within the more general approach including the Pomeron loops, the evolution Hamiltonian can be written in terms of the $n$-gluon "emission amplitudes"
 $Q_n$ \cite{YY}.
 
\beq\label{chi}
H[\rho,{\delta\over\delta\rho}]\,=\,\frac{d}{dY}\,\langle 0_a|\Omega^\dagger_Y(\rho,a)\,
\Omega_Y(R\,\rho, R\,a)|0_a\rangle{|_{Y=Y_0}}\,=\,\,\sum_n\,Q^\dagger_n[R]\,Q_n[R]\,.
\eeq
Each $Q_n$ depends on transverse coordinates and has a set of $n$ Lorentz and color indices which we do
 not indicate explicitly. Unfortunately the explicit form of $Q_n$ is not known at present.

Throughout this paper we will be thinking of the diagonal matrix element
$\langle v|\hat{\cal O}|v\rangle$ as being computed through the path integral representation
\beq
\langle v|\,\hat{\cal O}\,|v\rangle\,=\,\int D\rho\, \,\hat{\cal O}\,\,\,W[\rho]\,.
\eeq
As discussed in \cite{kl,kl5} this representation requires endowing $\rho(x)$ by additional ordering coordinate, but this subtlety is of no concern to us in this paper.

\subsection{High energy scattering}

Let us now consider the process whereby the projectile discussed above scatters on a target. The target is described by some distribution of target chromoelectric fields. 
We are working in the lightcone gauge natural to the 
projectile wave function, $A^-=0$. 
In this gauge the high energy scattering matrix of a single  gluon projectile at
 transverse position $x$ on the target is given by 
the eikonal factor
\begin{eqnarray}
\label{S}
S(x;0,x^-)\,\,=\,\,{\cal P}\,\exp\{i\int_{0}^{x^-} dy^-\,T^a\,\alpha^a_T(x,y^-)\}\,;
\,\,\,\,\,\,\,\,\,\,\,\,\,\,\,\,\,\,\,S(x)\,\equiv\,S(x;0,1)\,.
\end{eqnarray}

The field $\alpha_T$ is the large $A^+$ component created by the target color charge density. It
obeys the classical equation of motion  and is determined by the color charge density of the 
target $\rho_T(x)$ via\cite{first,JIMWLK,cgc} 
\beq\label{alpha}
\alpha^a_T(x,x^-)T^a\,\,=g^2\,\,{1\over \partial^2}(x-y)\,
\left\{S^\dagger(y;0,x^-)\,\,\rho^{a}_T(y,x^-)\,T^a\,\,S(y;0,x^-)\right\}\,.
\eeq
For a composite projectile which has some distribution of gluons in its wave function
 the eikonal $S$-matrix can be written in the form analogous to 
$S(x)$, see \cite{something}
\begin{equation}
\Sigma^{PP}[\alpha_T]\,\equiv\,\langle P|\,\hat S\,|P\rangle\,=\,\int d\rho_P\,\,W^P[\rho_P]\,\,
\,\exp\left\{i\int_{0}^{1} dy^-\int d^2x\,\rho_P^a(x,y^-)\,\alpha_T^a(x,y^-)\right\}
\label{s1}
\end{equation}
where $\hat S$ is the second quantized $S$-matrix operator of QCD. The quantity $\rho_P(x_i)$ is the color charge density in the projectile wave function at a given transverse position, while $W^P[\rho]$ can be thought of as the weight functional
which determines the probability density to find a given configuration of color charges in the projectile.
 For a single gluon $\rho^a(x_i)=T^a \delta^2(x_i-x^0_i)$, 
and eq.(\ref{s1}) reduces to eq.(\ref{S}). 

The total $S$-matrix of the scattering process at a given rapidity $Y$ is given by
\begin{equation}
{\cal S}(Y)\,=\,\int\, D\alpha_T^{a}\,\, W^T_{Y_0}[\alpha_T(x,x^-)]\,\,\Sigma^{PP}_{Y-Y_0}[\alpha_T(x,x^-)]\,.
\label{ss}
\end{equation}
In \eq{ss} we have chosen the frame where the target 
has rapidity $Y_0$ while the projectile carries the rest of the total rapidity $Y$.
Lorentz invariance  requires ${\cal S}$ to be independent of $Y_0$. 

The high energy evolution of the projectile wave function discussed in the previous subsection leads to the following expression for the evolution of the forward scattering matrix element:
\begin{equation}
\frac{d}{d\,Y}\,{\cal S}\,=-\,\int\, D\alpha_T^{a}\,\, W^T_{Y_0}[\alpha_T(x,x^-)]\,\,\,
H\left[\alpha_T,\frac{\delta}{\delta\,\alpha_T}\right]\,\,\,
\Sigma^{PP}_{Y-Y_0}[\alpha_T(x,x^-)]\,.
\label{hee}
\end{equation}
Here the (Hermitian) high energy evolution Hamiltonian $H$ can be viewed as acting
either to the right or to the left:
\begin{equation}
{\partial\over\partial Y}\,\Sigma^P\,\,=-\,\,H\left[\alpha_T,\,{\delta\over\delta\alpha_T}\right]\,
\,\Sigma^{PP}[\alpha_T]\,;
\ \ \ \ \ \ \ \ \ 
{\partial\over\partial Y}\,W^T\,\,=-\,\,
H\left[\alpha_T,\,{\delta\over\delta\alpha_T}\right]\,\,W^T[\alpha_T]\,.
\label{dsigma}
\end{equation} 
As was shown in \cite{something} in order for the total $S$-matrix to be Lorentz invariant and symmetric between 
the projectile and the target, the evolution Hamiltonian $H$ must be self dual. This is to say that the full evolution Hamiltonian which includes Pomeron loops has to be invariant under 
the Dense-Dilute Duality transformation
\begin{equation}\label{duality}
\alpha^a(x,x^-)\,\rightarrow \,i\,{\delta\over\delta\rho^a(x,x^-)},\ \ \ \ \ \ \ \ \ \ \ \ \ \ \ \, 
{\delta\over\delta\alpha^a(x,x^-)}\,\rightarrow\, -i\,\rho^a(x,x^-),\ \ \ \ \ \ \ \ \ \ \ \ \ \ \ \, S\,\rightarrow R
\end{equation}
However if one considers the situation where the target is large and the projectile is small (the dense-dilute limit), 
the symmetry between the target and the projectile is irrelevant. In the limit 
when the color charge density of the target is parametrically large ($\rho^a=O(1/\alpha_s)$) the Hamiltonian is given by
 the JIMWLK+ expression \cite{JIMWLK,cgc}
\begin{eqnarray}\label{JIMWLK}
H^{JIMWLK+}\,=\,\int_z\,\,  Q^a_i(z)\, Q^a_i(z)\,
\end{eqnarray}
with $Q[S]$ obtained from $\tilde Q[R]$  (\ref{Q}) 
by the substitution $R\rightarrow S$. In particular in the JIMWLK limit
\beq\label{qjim}
Q^a_i(z)\,=\,\int\, d^2x{(x-z)_i\over (x-z)^2}\,[S^{ab}(z)\,-\,S^{ab}(x)]\, J^b_R[S,x]\,.
\eeq
Thus for processes involving small perturbative projectile, the evolution of its wave function $W[\rho_P]$ is governed by $H^{KLWMIJ}$ while the evolution of the projectile averaged $S$-matrix $\Sigma[S]$ is given by  $H^{JIMWLK}$.

\subsection{Semi-inclusive reactions}

Consider an observable $\hat{\cal O}$ which depends only on the projectile degrees of freedom, which we want to calculate in the asymptotic state long time after the scattering has taken place. We assume that the target has been boosted to rapidity $Y_0$ and that the operator $\hat{\cal O}$ depends on degrees of freedom between the rapidity $Y_0+y$ and $Y$. As discussed in \cite{klw} such an observable is given by
\beq\label{O1}
\langle \hat{\cal O}\rangle\,=\,\langle T_{Y_0}\,\langle\,P_{Y-Y_0-y}|\,
 C^{\dagger}_{y}\,(1\,-\,\hat S^\dagger)\,
C_{y}\,\,\,\,\hat{\cal O}\,\,\,
C^{\dagger}_{y}\,
\,(1\,-\,\hat S)\,C_{y}\,
|P_{Y-Y_0-y}\,\rangle\,T_{Y_0} \rangle
\eeq
where $|T_{Y_0}\rangle$ is the wave function of the target boosted to rapidity $Y_0$ and  $|P_{Y-Y_0-y}\,\rangle$ is the wave function of the projectile boosted to rapidity $Y-Y_0-y$.
To calculate the evolution of this observable with respect to $y$ at fixed $Y_0$ and $Y-Y_0-y$, we consider first the same expression before taking its average over the target wave function
\beq\label{O2}
{\cal O}[S,\bar S]\,=\,\langle\,P_{Y-Y_0-y}|\,
 C^{\dagger}_{y}\,(1\,-\,\hat {S}^\dagger)\,
C_{y}\,\,\,\,\hat{\cal O}\,\,\,
C^{\dagger}_{y}\,
\,(1\,-\,\hat {\bar S})\,C_{y}\,
|P_{Y-Y_0-y}\,\rangle\,.
\eeq
The two are related by   
\beq
\langle \hat{\cal O}\rangle\,=\,\langle T_{Y_0}|\,{\cal O}[S,\bar S]|_{\bar S\,=\,S}\,|T_{Y_0} \rangle\,=\,
\int DS\,D\bar S\,W^t_{Y_0}[S]\,\delta(S\,-\,\bar S)\,\,{\cal O}[S,\bar S]\,.
\eeq
The derivation of the evolution is similar to that given in the first subsection:
\beq
{d {\cal O}[S,\bar S]\over dy}\,=\,
\lim_{\Delta y\rightarrow 0}{{\cal O}_{y+\Delta y}[S,\bar S]\,-\,
{\cal O}_{y}[S,\bar S]\over 
\Delta y}\,.
\eeq
We write $C_{y+\Delta y}\,=\,C_{\Delta y}\,C_y$ and expand $C_{\Delta y}$ up to quadratic order in $a$:
\begin{eqnarray}\label{expand}
C_{\Delta y}&=&\left\{1\,+\,i\,\int d^2x\,b_i^a(x)\int_{e^{y}\,\Lambda}^{e^{y+\Delta y}\,
\Lambda}{dk^+\over \pi^{1/2}| k^+|^{1/2}} 
\left[a^{ a}_i(k^+, x)\,+\,a^{\dagger a}_i(k^+, x))\right]\,-\,\right.\nonumber \\
&-&\left. \left(\int d^2x\,b_i^a(x)\int_{e^{y}\,\Lambda}^{e^{y+\Delta y}\,\Lambda}
{dk^+\over \pi^{1/2}| k^+|^{1/2}} 
\left[a^{ a}_i(k^+, x)\,+\,a^{\dagger a}_i(k^+, x))\right]\right )^2\right\}\,.
\end{eqnarray}
We now recall the identities derived in \cite{kl1}:
\begin{eqnarray}\label{iden}
&&\rho^a\,\hat {\bar S} \,| \psi\rangle\,=\,J_R^a[\bar S]\,\hat {\bar S}\,| \psi\rangle\,;
\ \ \ \  \ \ \ \ \ \ \ \
\hat {\bar S} \,\rho^a\,| \psi\rangle\,=\,J_L^a[\bar S]\,\hat {\bar S}\,| \psi\rangle\,;
\nonumber \\ 
&& \langle\psi |\,\rho^a\,\hat S^\dagger\,=\,J_L^a[S]\, \langle\psi |\,\hat S^\dagger\,;
\ \ \ \  \ \ \ \ \ \ \ \
\langle\psi |\,\hat S^\dagger\,\rho^a\,=\,J_R^a[S]\, \langle\psi |\,\hat S^\dagger\,.
\end{eqnarray}
Combining the expansion (\ref{expand}) and the identities (\ref{iden}) we obtain
\beq\label{evolo}
{d\over dy} {\cal O}[S,\bar S]\,=-\,H_3[S,\bar S]\,\,
 {\cal O}[S,\bar S]
\eeq
\beq H_3[S,\bar S]\,\equiv\, \int {d^2z\over 2\pi}\,
\left[Q^a_i(z,[S])\,+\,Q^a_i(z,[\bar S])\right]^2\,. 
\eeq
Let us introduce the notations
$$H_1[S]\,\equiv\,H^{JIMWLK}[S]\,,\ \ \ \ \ \ \ \ H_2[S,\bar S]\,\equiv\,H_1[S]\,+\,H_1[\bar S]\,.$$
The Hamiltonian $H_3$ was first introduced in \cite{HWS}.
It is straightforward to prove that the Hamiltonian $H_3$ obeys two useful identities.
First, when acting on a function $F$ which depends on the product  $S\,\bar S^\dagger$ only it reduces to the
action of the Hamiltonian $H_1$:
\beq\label{id1}
H_3[S,\bar S]\,\,F[S\,\bar S^\dagger]\,=\,H_1[S\,\bar S^\dagger]\,\,F[S\,\bar S^\dagger]\,.
\eeq 
Second, when $H_3$ acts on an arbitrary function $F[S,\bar S]$ and then $\bar S$ is set to be equal $S$
it is again equivalent to the action of $H_1$:
\beq\label{id2}
H_3[S,\bar S]\,\,F[S,\bar S^\dagger]|_{\bar S=S}\,=\,H_1[S]\,\,F[S, S^\dagger]\,.
\eeq
\eq{id2} can be written in the following form
\beq
\int D\bar S\,\delta(\bar S\,-\,S)\,H_3[S,\bar S]\,\,F[S,\bar S^\dagger]\,=\,H_1[S]\,\int D\bar S\,\delta(\bar S\,-\,S)
F[S,\bar S^\dagger]
\eeq
Referring back to eq.(\ref{evolo}) we see that for any observable  the evolution operator between rapidity $Y_1$ and $Y_2$ is
\beq\label{U3}
U_3({Y_1,\, Y_2})\,=\,\,\exp\left[-\,H_3(Y_2\,-\,Y_1)\right]\,.
\eeq

The preceeding discussion is easily generalized by considering a product of $n+1$ local in rapidity operators $\hat {\cal O}_i$ each one depending only on degrees of freedom at fixed rapidity $Y_0+Y_i$. We will call  $\hat {\cal O}_0$ (which plays a special role) the operator which depends only on the valence degrees of freedom of the projectile at rapidity $Y$.
The expectation value for such an observable reads
\bea\label{general}
\langle \hat {\cal O}_n \ldots \hat {\cal O}_0  \rangle&=&
\int DS\,D\bar S\,W^T_{Y_0}[S]\,\delta(S\,-\,\bar S)\,\times \nonumber \\
&\times& U_3({Y_0,\,Y_0+Y_n})\,{\cal O}_n[S,\bar S]\,
U_3({Y_n,\,Y_{n-1}})\,{\cal O}_{n-1}[S,\bar S]\,\ldots U_3({Y_0+Y_1,Y})\,{\cal O}_0[S,\bar S]\,.\nonumber \\
\eea
The operator $\hat O_0$ determines what kind of observation is made on the valence degrees of freedom of the projectile. In the present paper we are interested in observables which are inclusive over the valence projectile degrees of freedom. For these observables (see \cite{klw})
\beq\label{O11}
 {\cal O}_0\,[S,\bar S]\,=\,1\,-\,\Sigma^{PP}_0[S^\dagger]\,-\,\Sigma^{PP}_0[\bar S]\,+\,
\Sigma^{PP}_0[S^\dagger\,\bar S]
\eeq 
and this is the expression we use in the rest of this paper.
If on the other hand the process of interest is elastic over the projectile, we would have
\beq
{\cal O}_{0 \,{\rm elastic}}[S,\bar S] \,=\,(1\,-\,\Sigma^{PP}_0[S^\dagger])\,(1\,-\,\Sigma^{PP}_0[\bar S])\,.
\eeq
Finally for example jets produced in association with diffractive vector meson production at 
projectile`s rapidity one has $\hat{\cal O}_0=|\Psi\rangle\,\langle \Psi|$ with $\Psi$ be 
a wavefunction of a final vector meson state.

\section{Inclusive gluon production}

\subsection{Single inclusive gluon production}
Following the derivation of the previous section we define the operator $\hat{\cal O}_g$ corresponding to an inclusive production of a gluon with momentum $k_1$ 
at rapidity $Y_0+Y_1$
\beq\label{Og}
{\cal O}_g[S,\bar S]\,=\,\frac{d }{d\, y}\,\langle 0_a| C_{y}\,(1\,-\,\hat S^\dagger)\,
C_{y}^\dagger\,\,\,\hat n(k_1,y)\,\,\,
C_{y}\,(1\,-\,\hat {\bar S})\,C_{y}^\dagger|0_a\rangle|_{y=Y_0+Y_1}
\eeq
\beq
\hat n(k_1,y)\,=\,\int^{e^y\,\Lambda}_{e^{(Y_0+Y_1)}\,\Lambda}\, dk^+\,a^{\dagger\,a}_i(k_1,k^+)\,a^a_i(k_1,k^+)\,.
\eeq
In the dilute limit with the operator $C$ given in eq.(\ref{co}) the averaging over the state $|0_a\rangle$ can be performed explicitly. This requires expanding the exponential operator $C$ to second order in Taylor expansion. The result is
\beq\label{Og2}
{\cal O}_g(k_1)\,=\,
\int {d^2 z\over 2\,\pi}\,{d^2 \bar z\over 2\,\pi}\, \,e^{i\,k_1\,(z\,-\,\bar z)}\,
\,Q^a_i(z,[S])\,Q^a_i(\bar z,[\bar S])
\eeq
with the gluon production amplitude $Q$ the same as in eq.(\ref{qjim}).
Substituting \eq{Og2} into equation \eq{general} and remembering to set $\bar S=S$ in the
end of calculation we obtain for single inclusive gluon production
\begin{eqnarray}\label{Og3}
{d\sigma\over dY_1\,dk_1^2}\,&=&\, \langle \hat n(k_1,{Y_0+Y_1})\,\hat{\cal O}_0\rangle\,\\
&=&\,
\int DS\,D\bar S\,W^T_{Y_0}[S]\,\delta(\bar S\,-\,S)\,
U_3({Y_0,\,Y_0+Y_1})\,{\cal O}_g(k_1)\,U_3({Y_0+Y_1,Y})\,\Sigma^{PP}_{0}[S^\dagger \bar S]\nonumber
\end{eqnarray}
as only the last term in eq.(\ref{O11}) gives a nonvanishing contribution.
Using the identity \eq{id2} we obtain
\begin{eqnarray}\label{Og4}
{d\sigma\over dY_1\,dk_1^2}&=&
\int DS\,W^T_{Y_0}[S]\,U_1({Y_0,\,Y_0+Y_1})\, \int D\bar S\,\delta(\bar S\,-\,S)\,
 {\cal O}_g(k_1)\,U_3({Y_0+Y_1,Y})\, \Sigma^{PP}_{0}[S^\dagger \bar S]
 \nonumber \\
&=&\int DS\,D\bar S\,W^T_{Y_0+Y_1}[S]\,\delta(\bar S\,-\,S)\,
{\cal O}_g(k_1)\,U_3({Y_0+Y_1,Y})\, \Sigma^{PP}_{0}[S^\dagger \bar S]\,.
\end{eqnarray}
Using the fact that $\Sigma^{PP}$ evolves according to JIMWLK evolution we can write it as
\begin{eqnarray}\label{Og5}
{d\sigma\over dY_1\,dk_1^2}\,=\,\int DS\,D\bar S\,W^T_{Y_0+Y_1}[S]\,\delta(\bar S\,-\,S)\,
 {\cal O}_g(k_1)\,\Sigma^{PP}_{Y-Y_1-Y_0}[S^\dagger \bar S]\,.\,
\end{eqnarray}
Substituting eq.(\ref{Og2}) and assuming color neutrality of the target this becomes ($\bar Y_0\equiv Y_0+Y_1$)
\begin{eqnarray}
{d\sigma\over dY_1\,dk_1^2}&=&{\alpha_s\over 4\,\pi^3}\,
\int_{z,\bar z} e^{i\,k(z\,-\,\bar z)}\,\int_{x,y} {(z-x)_i\over (z-x)^2}\,
{(\bar z-y)_i\over (\bar z-y)^2}\,\int DS\, W^T_{\bar Y_0}[S]\,\,
{\rm tr} \left[(S_z^\dagger\,-\,S_x^\dagger)( S_{\bar z}\,-\, S_y)\right]\,\times\nonumber \\
&\times& \left[J_R^a[S,x]\,J_R^a[\bar S,y]\, \Sigma^{PP}_{Y-\bar Y_0}[S^\dagger \bar S]\right]_{\bar S=S}\,.
\end{eqnarray}
In order to make contact with the known results \cite{Kovchegov:2001sc, BKNW} and also to streamline the future discussion let us change variables from $\bar S$ to $U$ defined as:
\beq
\bar S\,=\,S\,U\, \ \ \ \  \ \ \ \ \ \ \ \ U\,=\,S^\dagger\,\bar S\,;\  \ \ \  \ \ \  \ \ \  \ \ \ \ U^\dagger \,U\,=\,1\,.
\eeq
In the end of the computation we will have to set $U=1$. 

It is straightforward to derive the following  identities  
\begin{eqnarray}
J^a_R[\bar S]|_{S,\bar S}&=&J^a_R[ U]|_{S,U}\,;\ \ \ \  \ \ \ \ \ \ \ \ \ \ \ \  \ \ \ \ \ \ \ \ \ \ \ \ \  \ \ \ \ \  J^a_L[\bar S]|_{S,\bar S}\,=\,S^{ab}\,J^b_L[ U]|_{S,U}\,;\nonumber \\
J^a_R[S]|_{S,\bar S}&=&J^a_R[ S]|_{S,U}\,+\,J^a_L[ U]|_{S,U};\ \ \ \  \ \ \ \ \ \ \ \  \ \ \ \  \ \  
J^a_L[ S]|_{S,\bar S}\,=\,J^a_L[ S]|_{S,U}\,+\,S^{ab}\,J^b_L[ U]|_{S,U}\,.
\end{eqnarray}
Using those we obtain
\beq
n^P(x,y;Y-\bar Y_0)\,\equiv\,
\left[J_R^a[S,x]\,J_R^a[\bar S,y]\, \Sigma^{PP}_{Y-\bar Y_0}[S^\dagger \bar S]\right]_{\bar S=S}\,=\,
\left[J_R^a[U,x]\,J_R^a[U,y]\, \Sigma^{PP}_{Y-\bar Y_0}[U]\right]_{U=1}\,.
\eeq
The function $n^P$ obeys the BFKL equation. To see this one can use  the identities (\ref{iden})
and express $n^P$ as a correlator of projectile color charge density operators 
\beq
n^P(x,y)\,=\,\int D\rho_P\, W^P[\rho_P]\,\,\rho^a_P(x)\,\rho^a_P(y)
\eeq
In the dilute limit $W^P$ evolves with the KLWMIJ Hamiltonian, which when integrated by parts to act on $\rho\,\rho$ reduces
to the BFKL equation. The final result for single gluon inclusive production reads
\begin{eqnarray}\label{Og31}
{d\sigma\over dy_1\,dk_1^2}&=&{\alpha_s\over \,\pi}\,
\int_{z,\bar z} e^{i\,k(z\,-\,\bar z)}\,\int_{x,y} {(z-x)_i\over (z-x)^2}\,
{(\bar z-y)_i\over (\bar z-y)^2}\,\times\,\nonumber \\
&\times & n^P(x,y;Y-\bar Y_0)\,\,[\langle T_{z,y}\rangle_{\bar Y_0}\,+\,\langle T_{x,\bar z}\rangle_{\bar Y_0}
\,-\,\langle T_{z,\bar z}\rangle_{\bar Y_0}\,-\,\langle T_{x,y}\rangle_{\bar Y_0}]
\end{eqnarray}
with  $\langle T \rangle$ denoting an $S$-matrix of a gluonic dipole:
\beq
\langle T_{x,y}\rangle_{\bar Y_0}\,\equiv\,\int DS\,W^t_{\bar Y_0}[S]\,{\rm tr}[S^\dagger_x\,S_y]\,.
\eeq  

\subsection{Multi-gluon production}

The formal generalization of single gluon inclusive cross section to an arbitrary number of gluons is straightforward.
We first introduce a generating functional ${\cal Z}$:
\beq\label{Z}
{\cal Z}[j]\,\equiv\,\int DS\, W^T_{Y_0}[S]\,\int D\bar S\,\delta(\bar S\,-\,S)\,\,
 {\cal P}\,e^{\,-\,\int_{Y_0}^{Y}\,d\eta\,\left[H_3\,-\,\int_k\, {\cal O}_g(k)\,j(\eta,k)\right]}\,\,\Sigma^{PP}_0[S^\dagger\bar S]
\eeq
The $n$-gluon cross section is calculated as \footnote{Note that instead of introducing the current $j$ we could introduce two different currents coupled 
separately to $Q[S]$ and $Q[\bar S]$. That would be in analogy with the Schwinger-Keldish formalism discussed
recently in \cite{francois}.}  
\beq\label{sign}
{d\sigma\over dY_1\,dk_1^2\,\ldots\, dY_n\,dk_n^2}\,=\,{\delta\over \delta j(Y_1,k_1)}\,\cdots\,
{\delta\over \delta j(Y_n,k_n)}\,\,\,{\cal Z}[j]\,|_{j=0}\,.
\eeq
It is interesting to note that the generating functional ${\cal Z}$ can be also used to define diffractive processes with
rapidity  gap(s). To impose a rapidity within an interval between $Y_1$ and $Y_2$ we simply set $j(\eta,k)= 2$ within
this interval and zero elsewhere. This procedure changes
the evolution Hamiltonian within this gap to $H_2$. As discussed in \cite{klw} this is precisely what is needed 
to forbid gluon emission within the rapidity gap.

The formal expression eq.(\ref{sign}) is unfortunately of little help as it stands. We can make a little bit of headway in understanding its structure by introducing the variable $U$ as in the previous subsection. A straightforward calculation allows one to express both the Hamiltonian and the gluon production operator as follows 
\beq\label{H3U}
H_3\,=\,H_1[S]\,+\,H_1[U]\,+\,V_{SU}[S,U]\,; \ \ \ \ \ \ \ \  \ \ \ \ \ \ \ 
V_{SU}[S,U]\,=\,2\,\int_z\,Q^a_z[S]\,Q^a_z[U]\,,
\eeq
\begin{eqnarray}
 {\cal O}_g(k)&=&\frac{\bar \alpha_s}{\pi}\,\int_{x,y,z,\bar z}\,e^{i\,k\,(z\,-\,\bar z)}\,{\cal N}_{x,z,y,\bar z}\,\left\{
J^a_L[x,U]
\left[(S_z^\dagger\,-\,S_x^\dagger)\,(S_{\bar z}\,U_{\bar z}\,-\,S_y\,U_y)\right]^{ab}\,
J^b_R[y,U]
\right.\,+ \nonumber \\
&&\ \ \ \ \ \ \ \ \ \  \ \ \ \ \ +\ \left.
J^a_R[x,S]
\left[(S_z^\dagger\,-\,S_x^\dagger)\,(S_{\bar z}\,U_{\bar z}\,-\,S_y\,U_y)\right]^{ab}\,
J^b_R[y,U]
\right\}
\end{eqnarray}
where the kernel ${\cal N}$ is defined in eq.(\ref{N}).

In the new variables the generating functional ${\cal Z}$ reads:
\beq\label{Z1}
{\cal Z}[j]\,=\,\int DS\,DU\, W^T_{Y_0}[S]\,\delta(U\,-\,1)\,\,
 {\cal P}\,e^{\,-\,\int_{Y_0}^{Y}\,d\eta\,\left[H_3\,-\,\int_k\, {\cal O}_g(k)\,j(\eta,k)\right]}\,\,\Sigma^{PP}_0[U]
\eeq
and the multigluon production cross section
\begin{eqnarray}\label{sign1}
{d\sigma\over dY_1\,dk_1^2\,\ldots\, dY_n\,dk_n^2}&=&
\,\int DS\,DU\, W^T_{Y_0+Y_1}[S]\,\delta(U\,-\,1)\,
 \,{\cal O}_g(k_1,[S,U])\,U_3(Y_2,\,Y_1) \,\times \nonumber\\ &\times& {\cal O}_g(k_2,[S,U])\,
 U_3(Y_3,\,Y_2)...
\,{\cal O}_g(k_n,[S,U])\,\Sigma^{PP}_{Y-Y_n-Y_0}[U]\,.
\end{eqnarray}
Here we have used the fact that both, the constant function and the delta function are eigenfunctions of the operators $Q^a_i[x,S]$ with vanishing eigenvalue. That allowed us to drop the $U$-dependent terms in leftmost evolution operator and the $S$-dependent terms in the rightmost evolution operator, and thus simply to evolve $W^T[S]$ and $\Sigma^{PP}[U]$ to the rapidity of the closest emitted gluon. A further slight simplification that follows from the same argument is that in ${\cal O}_g(k_n)$ we can drop the second term which involves $J_L[S]$, while in ${\cal  O}_g(k_1)$ we can set $U=1$;
\beq\label{righto}
{\cal O}_g(k_n)\rightarrow\frac{\bar \alpha_s}{\pi}\,\int_{x,y,z,\bar z}\,e^{i\,k_n\,(z\,-\,\bar z)}\,{\cal N}_{x,z,y,\bar z}\,\left\{
J^a_L[x,U]
\left[(S_z^\dagger\,-\,S_x^\dagger)\,(S_{\bar z}\,U_{\bar z}\,-\,S_y\,U_y)\right]^{ab}\,
J^b_R[y,U]
\right\}
\eeq
\begin{eqnarray}\label{lefto}
{\cal O}_g(k_1)&\rightarrow&\frac{\bar \alpha_s}{\pi}\,\int_{x,y,z,\bar z}\,e^{i\,k_1\,(z\,-\,\bar z)}\,{\cal N}_{x,z,y,\bar z}\,\left\{
J^a_L[x,U]
\left[(S_z^\dagger\,-\,S_x^\dagger)\,(S_{\bar z}\,-\,S_y\,\right]^{ab}\,
J^b_R[y,U]
\right.\,+ \nonumber \\
&&\ \ \ \ \ \ \ \ \ \  \ \ \ \ \ +\ \left.
J^a_R[x,S]
\left[(S_z^\dagger\,-\,S_x^\dagger)\,(S_{\bar z}\,-\,S_y\,\right]^{ab}\,
J^b_R[y,U]
\right\}\,.
\end{eqnarray}
We note the structure of eq.(\ref{sign1}) is rather curious from the point of view of the Hilbert space of the JIMWLK Hamiltonian \cite{YY}. The expression involves the off diagonal matrix element of the evolution operator with insertions of ${\cal O}_g$'s between two states which are products of factors depending on $S$ and $U$: $ W^T_{Y_0+Y_1}[S]\,\delta(U\,-\,1)$ and $\Sigma^{PP}_{Y-Y_n-Y_0}[U]$. The factor $\delta(U\,-\,1)$ on the left is nothing but the Yang ground state of $H^{JIMWLK}[U]$, while the unity on the right is the Yin ground state of $H^{JIMWLK}[S]$ \cite{YY}. Thus the expression for the multigluon production incorporates some elements of the Dense-Dilute Duality symmetry. This can possibly give a clue for future progress in calculating multigluon production.

At the moment however despite the simplification of eqs.(\ref{righto},\ref{lefto}) the calculation of multigluon production within the general JIMWLK looks like a very difficult problem. The Hamiltonian $H_3$ couples $S$ and $U$ in a nonperturbative way and this coupled problem has to be solved. We note that since the evolution of $U$ is eventually projected onto the Yang state, one can develop perturbation theory for the $U$ part of the evolution. In particular writing
\beq 
U(x)\,=\,\exp\{i\,\beta^a\,T^a\} 
\eeq
and considering $\beta$ to be small we can perform expansion in powers of $\beta$. This is useful since the Yang projection sets $\beta$ to be equal to zero at the end of the calculation. As we shall see in the next section this strategy is very useful within the dipole model approximation. Here however, even if $\beta$ is considered to be small, its interaction with $S$ in $H_3$ is not suppressed. Thus the first step in this perturbation theory requires 
solution of the coupled $S$ and $\beta$ theory. This problem is beyond our abilities at the moment. We can make however quite a bit of headway by restricting to the dipole limit. This we do in the next section in the hope to come back to the general problem in the future.

\section{Multi-gluon production in the dipole limit}
In this section we discuss the dipole limit of the formulae derived in the previous section. We will see that in this limit the multigluon production amplitudes simplify dramatically. The starting point of our discussion is eq. (\ref{Z}).
\subsection{The dipole limit}

The dipole operator for $S$ and $\bar S$ are  defined conventionally as 
 \beq
s_{x,y}\,=\,\frac{1}{N}\,tr[S_F(x)\,S^\dagger_F(y)]\,;\ \ \ \ \ \ \ \ \ \ \ \ \ \
\bar s_{x,y}\,=\,\frac{1}{N}\,tr[\bar S_F(x)\,\bar S^\dagger_F(y)]\,.
\eeq 
We will also have to consider quadrupole operators 
\beq
q_{x,y,u,v}\,=\,\frac{1}{N}\,tr[S_F(x)\,S^\dagger_F(y)\,S_F(u)\,S^\dagger_F(v)]\,;\ \ \ \ \ \ \ \ \ \ \ \ \ \
\bar q_{x,y,u,v}\,=\,\frac{1}{N}\,tr[\bar S_F(x)\,\bar S^\dagger_F(y)\,\bar S_F(u)\,\bar S^\dagger_F(v)]\,.
\eeq
In the discussion of the total cross section the quadrupole operator appears in $1/N$ suppressed terms, however in less inclusive observables \cite{klw} it can contribute already in the leading order. This will be the case in our present discussion as well.

To define the dipole model limit we assume that the initial  state of the projectile $|P\rangle$ contains only dipoles.  We also make two important approximations. First, we consider only the leading $N$ part of the evolution.
This drastically simplifies the calculations since in this limit there are no transitions between different singlet states during scattering. Technically this means \cite{klw} that any singlet operator ($s$ or $q$ etc.) satisfies a first order differential equation in rapidity. Second, we assume the target mean field approximation in which
the target averages for dipoles and quadrupoles factorize.
\begin{eqnarray}
&&\langle s(x,y)\, s(u,v)\rangle_T\,=\, \langle s(x,y)\rangle_T\,\langle s(u,v)\rangle_T; \\
&&\langle q(x,y,u,v)\, s(z,\bar z)\rangle_T\,=\, \langle q(x,y,u,v)\rangle_T\,\langle s(z,\bar z)\rangle_T\nonumber \\
&&\langle q(x,y,u,v)\, q(\bar x,\bar y, \bar u,\bar v)\rangle_T\,=\, \langle q(x,y,u,v)\rangle_T\,
\langle q(\bar x, \bar y, \bar u,\bar v)\rangle_T\nonumber
\end{eqnarray}
Physically this approximation is tantamount to assuming no correlations between the target fields. 
This is the same approximation as employed in deriving the Kovchegov equation \cite{Kovchegov}.

\subsubsection{The Hamiltonian $H_3$}

The action of the JIMWLK Hamiltonian on a function $F$ which depends only on $s$ and $q$  in the large $N$ limit
can be straightforwardly written as:
\beq
H^{JIMWLK}[S]\,F[s,q]\,=\,[H_s\,+\,H_q]\,F[s,q]
\eeq
with
\begin{eqnarray}\label{sq}
H_s &=& -\frac{\bar \alpha_s}{2\,\pi}\,\int_{x,y,z} M_{x,y;z}\,[s_{x,z}\,s_{z,y}\,-\,s_{x,y}]\,{\delta\over \delta s(x,y)}\,;
\nonumber \\
H_q &=& -\frac{\bar \alpha_s}{2\,\pi}\,\int_{x,y,u,v,z} \left\{-\,[M_{x,y;z}\,+\,M_{u,v;z}\,-\,L_{x,u,v,y;z}]
\,\,q_{x,y,u,v}\,-\,
 L_{x,y,u,v;z}\,s_{x,v}\,s_{y,u}\,-\,L_{x,v,u,y;z}\,s_{x,y}\,s_{u,v}\right.\,\nonumber \\
&+&\,L_{x,v,u,v;z}\,q_{x,y,u,z}\,s_{z,v}
\left.+\,L_{x,y,x,v;z}\,q_{z,y,u,v}\,s_{x,z}
\,+\,L_{x,y,u,y;z}\,q_{x,z,u,v}\,s_{z,y}\,+\,L_{u,y,u,v;z}\,q_{x,y,z,v}\,s_{u,z}\right\}\,\nonumber \\
&& \ \ \ \ \ \ \ \ \ \ \ \ \ \times \ {\delta\over \delta q(x,y,u,v)}\,.
\end{eqnarray}
Here in addition to the Weiszacker-Williams kernel
\beq
K_{x,y;z}\,=\,{(x\,-\,z)_i\,(y\,-\,z)_i\over (x\,-\,z)^2\,(y\,-\,z)^2}
\eeq
we have introduced the dipole kernel
\beq
M_{x,y;z}\,=\,K_{x,x,z}\,+\,K_{y,y,z}\,-\,K_{x,y,z}\,-\,K_{y,x,z}\,=\,
{(x\,-\,y)^2\over (x\,-\,z)^2\,(y\,-\,z)^2}
\eeq
and the "quadrupole kernel" \cite{MC,kl1}
\beq
L_{x,y,u,v;z}\,=\,\left[{(x\,-\,z)_i\over (x\,-\,z)^2}\,-\,{(y\,-\,z)_i\over (y\,-\,z)^2}\right ]\,\,
\left[{(u\,-\,z)_i\over (u\,-\,z)^2}\,-\,{(v\,-\,z)_i\over (v\,-\,z)^2}\right ]
\eeq
The kernel $L$ can be expressed in terms of the kernel $M$ as
\beq
L_{x,y,u,v;z}\,=\,\,{1\over 2}\,\left[\,M_{y,u;z}\,+\,M_{x,v;z}\,-\,M_{y,v;z}-M_{x,u;z}\,\right]\,.
\eeq
For future use we introduce an additional kernel
\beq\label{N}
{\cal N}_{x,z,u,\bar z}\,=\,{(x-z)_i\,(u-\bar z)_i\over (x-z)^2\,(u-\bar z)^2}\,.
\eeq
The equation for $s$ is the Kovchegov equation \cite{Kovchegov}, while the equation for $q$ has been derived in \cite{JMK, klw}.

The fact that the equation is first order means that the solution for the evolution of any function of $s$ and $q$ which satisfies 
the JIMWLK evolution in the dipole limit is \cite{LL1,LL2}
\beq
F_Y[s,q]\,=\,F_{Y=0}[s_Y,\, q_Y]
\eeq
with $s_Y$ and $q_Y$ solutions of the differential equations
\beq\label{dipoleeq}
\partial_\eta\,s(x,y)=\frac{\bar \alpha_s}{2\,\pi}\,\int_{x,y,z} M_{x,y;z}\,[s_{x,z}\,s_{z,y}\,-\,s_{x,y}]
\eeq
and
\begin{eqnarray}\label{quadrupoleeq}
\partial_\eta\,q(x,y,u,v)&=&\frac{\bar \alpha_s}{2\,\pi}\,\int_{x,y,u,v,z} -\,[M_{x,y;z}\,+\,M_{u,v;z}\,-\,L_{x,u,v,y;z}]
\,\,q_{x,y,u,v}\,\nonumber\\
&-&\, L_{x,y,u,v;z}\,s_{x,v}\,s_{y,u}\,-\,L_{x,v,u,y;z}\,s_{x,y}\,s_{u,v}
+\,L_{x,v,u,v;z}\,q_{x,y,u,z}\,s_{z,v}\nonumber\\
&+&\,L_{x,y,x,v;z}\,q_{z,y,u,v}\,s_{x,z}
\,+\,L_{x,y,u,y;z}\,q_{x,z,u,v}\,s_{z,y}\,+\,L_{u,y,u,v;z}\,q_{x,y,z,v}\,s_{u,z}
\end{eqnarray}
with the initial conditions 
\beq
s_{Y=0}(x,y)\,=\,s(x,y),\ \ \ \ \ \ \ \ \ \ \ \ \ \ \ \ \ \ \ \ \ \ \ \ q_{Y=0}(x,y,u,v)\,=\,q(x,y,u,v)\,.
\eeq
The quadrupoles and dipoles are not quite independent, in the sense that 
\beq\label{constr}
q(x,y,y,v)\,=\,s(x,v)\,,\ \ \ \ \ \ \ \ \ \ \ \ \ \ \ \ \ \ \ q(x,y,u,x)\,=\,s(u,y)\,.
\eeq
It can be checked that eqs. (\ref{dipoleeq},\ref{quadrupoleeq}) preserve this constraint. That is if the initial condition at 
$\eta=0$ satisfies eq.(\ref{constr}) so does the solution of eqs.(\ref{dipoleeq},\ref{quadrupoleeq}) at any rapidity $\eta$.

The dipoles $s$ and the quadrupoles $q$ do not exhaust all color singlet operators that enter our analysis. We also have to consider quadrupoles of the mixed type
\beq
q_{x,y,v,u}^{s\bar s}\,=\,\frac{1}{N}\,tr[S_F(x)\,S^\dagger_F(y)\,\bar S_F(u)\,\bar S^\dagger_F(v)]\,.
\eeq
These objects appear at the intermediate stages of the calculation even though at the end 
of  computations (\eq{Z1}) we always set $\bar S=S$. 
The reason this operator appears is that the projectile averaged scattering matrix $\Sigma^{PP}_0$ in eq.(\ref{Og4},\ref{Og5}) depends on the product $S(x)\bar S^\dagger(x)$. The dipole content of the projectile wave function then means that $\Sigma^{PP}_0$ 
depends on the dipole operator built from $S(x)\bar S^\dagger(x)$ which is just a special case of $q^{s\bar s}$.

 For the
mixed quadrupole $q^{s\bar s}$ 
\bea\label{qss1}
&&q_{x,y,u,v}^{s\bar s}|_{\bar S=S}\,=\,q_{x,y,u,v}\,; \ \ \ \ \ \ \ \ \  \ \ \ \ 
q_{x,y,y,v}|_{\bar S=S}\,=\,s_{x,v}\,; \ \ \ \  \ \ \ \ \ \  \ \ \ \  \ \ \ \ \ \ \  \ q_{x,y,y,x}|_{\bar S=S}=1\,; \nonumber \\
&&q_{x,x,u,v}^{s\bar s}\,=\, \bar s_{u,v}\,; \ \ \ \ \ \ \ \ \   \ \ \ \ \ \ \  \ \ \ \ \ \ \ q_{x,y,u,u}^{s\bar s}\,=\, s_{x,y}\,.
\eea 
The action of the JIMWLK Hamiltonian on a function $F$ which depends on  $q^{s\bar s}$ in addition to $s$ and $q$ in the large $N$ limit
can be written as:
\beq
H^{JIMWLK}[S]\,F[s,q, q^{s\bar s}]\,=\,\left(H_s\,+\,H_q\,+\,H_{q^{s\bar s}}\right)\,F[s,q,q^{s\bar s}]
\eeq
with
\begin{eqnarray}
H_{q^{s\bar s}} &=& -\frac{\bar \alpha_s}{2\,\pi}\,\int_{x,y,u,v,z} \left\{-\,[M_{x,y;z}\,+\,K_{x,y;z}]
\,\,q^{s\bar s}_{x,y,u,v}\,+\,K_{x,y;z}\,s_{x,y}\,s_{u,v}\,+\,\right.\nonumber \\ 
&+&\,\left.(K_{x,x;z}\,-\,K_{x,y;z})\,q^{s\bar s}_{z,y,u,v}\,s_{x,z}
\,+\,(K_{y,y;z}\,-\,K_{x,y;z})\,q^{s\bar s}_{x,z,u,v}\,s_{z,y}\right\}\,\ {\delta\over \delta q^{s\bar s}(x,y,u,v)}\,.
\end{eqnarray}

We also have to derive the dipole limit of the operator $Q_i^a[z,S]\,Q_i^a[\bar z,\bar S]$ which enters both
the evolution Hamiltonian $H_3$ and the expression for the gluon emission vertex ${\cal O}_g$. 
Again using the large $N$ limits to evolve color singlets we obtain
\begin{eqnarray}\label{operd}
&& 2\,Q_i^a[z,S]\,Q_i^a[\bar z,\bar S]\,=\,\frac{\bar \alpha_s}{2\,\pi}\,\int_{x,y,v,u}\,\left\{
-\,({\cal N}_{x,z,v,\bar z}\,+\,{\cal N}_{y,z,u,\bar z})\,q^{s\bar s}_{x,y,u,v}\,+\right.\nonumber \\
&&\ \ \ \ \ \ \ +\
({\cal N}_{y,z,v,\bar z} +\ {\cal N}_{x,z,u,\bar z})\,s_{x,y}\,\bar s_{u,v} +\ 
({\cal N}_{y,z,v,\bar z}\,+\,{\cal N}_{x,z,u,\bar z}\,-\,{\cal N}_{x,z,v,\bar z}\,-\,{\cal N}_{y,z,u,\bar z})\,
q^{s\bar s}_{z,y,u,\bar z}\,q^{s\bar s}_{x, z,\bar z, v}\,\nonumber \\
&&\ \ \ \ \ \ \ +\ 
({\cal N}_{x,z,v,\bar z}\,-\,{\cal N}_{y,z,v,\bar z})\,s_{x,z}\,q^{s\bar s}_{z,y,u,v}\,
+\,({\cal N}_{y,z,u,\bar z}\,-\,{\cal N}_{x,z,u,\bar z})\,s_{z,y}\,q^{s\bar s}_{x,z,u,v}\,+\nonumber \\
&& \ \ \ \ \ \ \  + \ \left.
({\cal N}_{x,z,v,\bar z}\,-\,{\cal N}_{x,z,u,\bar z})\,\bar s_{\bar z,v}\,q^{s\bar s}_{x,y,u,\bar z}\,
+\,({\cal N}_{y,z,u,\bar z}\,-\,{\cal N}_{y,z,v,\bar z})\,\bar s_{u,\bar z}\,q^{s\bar s}_{x,y,\bar z,v}
\right\}\,\,{\delta\over \delta q^{s\bar s}_{x,y,u,v}}\,.
\end{eqnarray}

Note a very important property of eq.(\ref{operd}): when acting on a function of $s, \ q$ and $q^{s\bar s}$ the operator does not produce any additional objects. In principle one might have expected that the operator involves higher mixed multipole operators. If that were the case we could not have limited ourselves by considering only the set of operators $s,\ q$ and $q^{s\bar s}$ but would have to introduce higher miltipoles. However as it stands taking into account that $\Sigma^{PP}_0$
 at initial rapidity depends only on $q^{s\bar s}$ we need not include additional operators into consideration at all.

Our  goal is to use the fact that we have to put $S=\bar S$ at the end of the calculation to the utmost and thereby to minimize the number of degrees of freedom which we have to deal with. To this end we note that the operator $ {\cal O}_g$
does not involve derivatives with respect to $s$, $q $ and $\bar s$, $\bar q$ but only with respect to $q^{s\bar s}$. This allows us to simplify the evolution operator $H_3$ by using
\beq\label{simp}
\left[\,\left(H_s\,+\,H_{\bar s}\,+\,H_q\,+\,H_{\bar q}\right)\,F[s,q;\bar s,\bar q]\,\right]_{\bar s=s;\, \bar q=q}
\,=\,\left(H_s\,+\,H_q\right)\,F[s,q;s,q]
\eeq 
and  substitute $s$ ($q$) instead of  $\bar s$ ($\bar q$)
in expressions for $ {\cal O}_g$  and  $H_{q^{s\bar s}}$. 

Since $ {\cal O}_g$ depends on derivative with respect to $q^{s\bar s}$ we cannot set $q^{s\bar s}$ equal $q$ from the outset.
We therefore define the "fluctuation" of $q^{s\bar s}$ by
\beq\label{shift}
q_{x,y,v,u}^{s\bar s}\,=\,q_{x,y,v,u}\,+\,t_{x,y,v,u}\,; \ \ \ \ \ \ \ \ \ \ \ \ \ \ \ \ \ \ 
{\delta \over \delta q_{x,y,v,u}^{s\bar s}}\,=\,{\delta \over \delta t_{x,y,v,u}}\,.
\eeq
As defined above 
\beq
t_{x,x,v,u}\,=\,t_{x,y,u,u}\,=\,0\,.
\eeq
This follows from \eq{qss1}.

At the end of the computation we have to  set $t_{x,y,v,u}=0$. We now change variables from $q^{s\bar s}$ to $t$ throughout. Note that $H_q$ as defined in 
\eq{sq} is proportional to $\delta/\delta q$ at fixed $q^{s\bar s}$. In order to correctly account for the shift
we have to remember that
\beq\label{pd}
{\delta\over \delta q}|_{q^{s\bar s}}\,=\,{\delta\over \delta q}|_{t}\,-\,{\delta\over \delta t}|_{q}\,.
\eeq

Now we have to express $\cal Z$ in terms of the dipole $s$, quadrupole $q$, and $t$. 
The $H_3$ Hamiltonian becomes
\beq\label{H3d}
H_3\,=\,H_s\,+\,H_q\,+\,H_t
\eeq
with
\begin{eqnarray}\label{Ht}
H_t&=-&\frac{\bar \alpha_s}{2\,\pi}\,\int_{x,y,u,v,z} \left\{-\,[M_{x,y;z}\,+\,M_{u,v;z}\,-\,L_{x,u,v,y;z}]
\,\,t_{x,y,u,v}\,+\,L_{x,v,u,v;z}\,t_{x,y,u,z}\,s_{z,v}\,+
 \right.\,\nonumber \\
&+&
 L_{x,y,x,v;z}\,t_{z,y,u,v}\,s_{x,z}
\,+\,L_{x,y,u,y;z}\,t_{x,z,u,v}\,s_{z,y}\,+\,L_{u,y,u,v;z}\,t_{x,y,z,v}\,s_{u,z} \nonumber \\
&-&\left.
L_{x,y,u,v;z}\,(t_{z,y,u,z}\,s_{x,v}\,+\,t_{x,z,z,v}\,s_{u,y})-\,L_{x,y,u,v;z}t_{z,y,u, z}\,t_{x, z,  z, v}
\right\}
 \ {\delta\over \delta t_{x,y,u,v}}\,.
\end{eqnarray}
Thus the four point function $t$ satisfies the classical equation
\begin{eqnarray}
{d\over d\eta}t_{x,y,u,v}&=&\frac{\bar \alpha_s}{2\,\pi}\,\int_{z} \left\{-\,[M_{x,y;z}\,+\,M_{u,v;z}\,-\,L_{x,u,v,y;z}]
\,\,t_{x,y,u,v}\,+\,L_{x,v,u,v;z}\,t_{x,y,u,z}\,s_{z,v}\,+
 \right.\,\nonumber \\
&+&
 L_{x,y,x,v;z}\,t_{z,y,u,v}\,s_{x,z}
\,+\,L_{x,y,u,y;z}\,t_{x,z,u,v}\,s_{z,y}\,+\,L_{u,y,u,v;z}\,t_{x,y,z,v}\,s_{u,z} \nonumber \\
&-&\left.
L_{x,y,u,v;z}\,(t_{z,y,u,z}\,s_{x,v}\,+\,t_{x,z,z,v}\,s_{u,y})-\,L_{x,y,u,v;z}t_{z,y,u, z}\,t_{x, z,  z, v} 
\right\}\,.
 \end{eqnarray}
The Hamiltonian $H_t$ describes propagation of $t$ in the background field $s$.  

The Hamiltonian eq.(\ref{Ht}) has the following very useful property.
Let us introduce the equal point limits of the function $t$:
the two-point function $\phi_{x,y}\,\equiv\,t_{x,y,y,x}$, and the three point function $\psi_{x,y,v}\equiv\,t_{x,y,y,v}$. The functions $\phi$ and $\psi$ are not fully independent and their mutual relation is precisely that of the
dipole $s$ and quadrupole $q$, $\psi_{x,y,x}=\phi_{x,y}$. Also $\phi_{x,x}=\psi_{x,x,v}=0$ as follows from \eq{qss1}.
The evolution of $\phi$  can be found by taking the  $x=v,\ y=u$ in eq.(\ref{Ht}) and similarly for $\psi$. The equations turns out to be decoupled from  $t_{x,y,u,v}$ with all four coordinates different
\beq\label{hamphi}
\partial_\eta\,\phi_{x,y}\,=\,\frac{\bar \alpha_s}{2\,\pi}\,\int_{z} M_{x,y;z}\,\,
[\phi_{x,z}\,+\,\phi_{z,y}\,-\, \phi_{x,y}+\phi_{z,y}\,\phi_{x, z}]
\eeq
\begin{eqnarray}\label{hampsi}
\partial_\eta\,\psi_{x,y,v}&=&
\frac{\bar \alpha_s}{2\,\pi}\,\int_{z} \Big\{-\,{1\over 2}[M_{x,y;z}\,+\,M_{y,v;z}\,+\,M_{x,v;z}]
\,\,\psi_{x,y,v}\,+\,L_{x,v,y,v;z}\,\psi_{x,y,z}\,s_{z,v}\,+
 L_{x,y,x,v;z}\,\psi_{z,y,v}\,s_{x,z}\nonumber\\
&&\ \ \ \ \ \ \ \ \ \ \ \ \  \ \ \ \ \ \ \ \ \ -\,L_{x,y,y,v;z}\left[\left(\phi_{z,y}+1\right)\,\psi_{x, z, v} -\phi_{z,y}\,s_{x,v}\right]\Big\}\,.
\end{eqnarray}
The equation for $\phi$ is equivalent to the Kovchegov equation for $1+\phi$. 
We will come back to these equations later.

The coupling of $t$ to $s$ is the origin of the violation of the
AGK cutting rule  found in \cite{JMK}.
In the following we will be expanding the evolution operator $\exp \{-H_3 Y\}$ in powers of $t$ since $t$ has to be set to zero at the end of the calculation. This is obviously tantamount to perturbation theory in the last "splitting" term in eq. (\ref{Ht}).
The highest order of the perturbation theory up to which we have to expand is determined 
by the number of emitted gluons.
But first we have to rewrite the gluon emission operator eq.(\ref{operd}) in terms of $t$.

\subsubsection{The gluon emission operator ${\cal O}_g$ }
The expression for the gluon emission operator is easily obtain, although it is cumbersome to write down. First we define for simplicity
\beq\label{Aop}
{\cal O}_g(k)=\frac{\bar \alpha_s}{2\,\pi}\,
\int_{z,\bar z} \,e^{i\,k\,(z\,-\,\bar z)}\,A(z,\, \bar z)
\eeq
The operator $A$ is linear in functional derivative $\delta/\delta t$. Counting the functional derivative as a power $t^{-1}$ the operator contains terms of order minus one, zero and  one in $t$. 
\beq
A(z,\, \bar z)=A_{-1}(z,\, \bar z)+A_0(z,\, \bar z)+A_1(z,\, \bar z)\label{ai}
\eeq
with
\begin{eqnarray}\label{ais}
A_{-1}&\equiv&\int_{x,y,u,v}\ J(x,y,u,v;z,\bar z)\  {\delta\over\delta t_{x,y,u,v}}\nonumber\\
&=&\int_{x,y,u,v}\left\{({\cal N}_{y,z,v,\bar z} +\ {\cal N}_{x,z,u,\bar z})\,s_{x,y}\, s_{u,v}-({\cal N}_{x,z,v,\bar z}\,+\,{\cal N}_{y,z,u,\bar z})\,q_{x,y,u,v}\right.\nonumber\\
&+&
({\cal N}_{x,z,v,\bar z}\,-\,{\cal N}_{y,z,v,\bar z})\,s_{x,z}\,q_{z,y,u,v}\,+\,
({\cal N}_{y,z,u,\bar z}\,-\,{\cal N}_{x,z,u,\bar z})\,s_{z,y}\,q_{x,z,u,v}\,\nonumber \\
&+&
({\cal N}_{x,z,v,\bar z}\,-\,{\cal N}_{x,z,u,\bar z})\,s_{\bar z,v}\,q_{x,y,u,\bar z}\,+\,
({\cal N}_{y,z,u,\bar z}\,-\,{\cal N}_{y,z,v,\bar z})\, s_{u,\bar z}\,q_{x,y,\bar z,v}
\nonumber \\ 
&+&\left.({\cal N}_{y,z,v,\bar z}\,+\,{\cal N}_{x,z,u,\bar z}\,-\,{\cal N}_{x,z,v,\bar z}\,-\,{\cal N}_{y,z,u,\bar z})\,
q_{z,y,u,\bar z}\,q_{x, z,\bar z, v}\right\} {\delta\over\delta t_{x,y,u,v}}\\
A_0&\equiv&\int_{x,y,v,u;\bar x,\bar y,\bar u,\bar v;z,\bar z}t_{\bar x,\bar y,\bar u,\bar v} \ 
D(x,y,v,u;\bar x,\bar y,\bar u,\bar v;z,\bar z)\ {\delta\over\delta t_{x,y,u,v}}\nonumber\\\label{aiss}
&=&\int_{x,y,v,u}\,\left\{
-\,({\cal N}_{x,z,v,\bar z}\,+\,{\cal N}_{y,z,u,\bar z})\,t_{x,y,u,v}\,\right.\nonumber \\
&+&({\cal N}_{y,z,v,\bar z}\,+\,{\cal N}_{x,z,u,\bar z}\,-\,{\cal N}_{x,z,v,\bar z}\,-\,{\cal N}_{y,z,u,\bar z})\,
q_{z,y,u,\bar z}\,t_{x, z,\bar z, v}\,\nonumber \\
&+&({\cal N}_{y,z,v,\bar z}\,+\,{\cal N}_{x,z,u,\bar z}\,-\,{\cal N}_{x,z,v,\bar z}\,-\,{\cal N}_{y,z,u,\bar z})\,q_{x, z,\bar z, v}\,
t_{z,y,u,\bar z}\,\,\nonumber \\
&+&\ 
({\cal N}_{x,z,v,\bar z}\,-\,{\cal N}_{y,z,v,\bar z})\,s_{x,z}\,t_{z,y,u,v}\,
+\,({\cal N}_{y,z,u,\bar z}\,-\,{\cal N}_{x,z,u,\bar z})\,s_{z,y}\,t_{x,z,u,v}\,\nonumber \\
&+& \ \left.
({\cal N}_{x,z,v,\bar z}\,-\,{\cal N}_{x,z,u,\bar z})\,s_{\bar z,v}\,t_{x,y,u,\bar z}\,
+\,({\cal N}_{y,z,u,\bar z}\,-\,{\cal N}_{y,z,v,\bar z})\, s_{u,\bar z}\,t_{x,y,\bar z,v}
\right\}\,\,{\delta\over \delta t_{x,y,u,v}}\\
A_1&\equiv&\int_{x,y,u,v}\ V(x,y,u,v;z,\bar z)\ t_{z,y,u,\bar z}\,t_{x, z,\bar z, v}\ {\delta\over \delta t_{x,y,u,v}}\nonumber\\ \label{aisss}
&=&\int_{x,y,u,v}\{({\cal N}_{y,z,v,\bar z}\,+\,{\cal N}_{x,z,u,\bar z}\,-\,{\cal N}_{x,z,v,\bar z}\,-\,{\cal N}_{y,z,u,\bar z})\,
t_{z,y,u,\bar z}\,t_{x, z,\bar z, v}\}{\delta\over \delta t_{x,y,u,v}}
\end{eqnarray}

The explicit form of $D$ and $J$ is given in Appendix A.  
It is also convenient for future use to define the "vertices" associated with the two point function $\phi_{x,y}$
\begin{eqnarray}\label{jphi}
J_\phi(x,y;z,\bar z)&\equiv& J(x,y,y,x;z,\bar z)\\ &=&
\left\{
{\cal N}_{y,z,x,\bar z}\,(s^2_{z,\bar z}\,+\,s^2_{x,y}\,-\,s^2_{x,z}\,-\, s^2_{y,\bar z})+{\cal N}_{y,z,y,\bar z}\,(s^2_{y,\bar z}\,+\,s^2_{z,y}\,-\,s^2_{z,\bar z}\,-\,1)\right. \,+\nonumber \\
&+&{\cal N}_{x,z,y,\bar z}\,(s^2_{z,\bar z}\,+\,s^2_{x,y}\,-\,s^2_{z,y}\,-\,\,s^2_{\bar z,x})\,
+\,{\cal N}_{x,z,x,\bar z}\,(s^2_{\bar z,x}\,+\,s^2_{x,z}\,-\,s^2_{z,\bar z}\,-\,1)\,\left.\right\}\,.\nonumber
\end{eqnarray}
We also separate the part of the vertex $D$ which corresponds to the first line of \eq{aiss} (the virtual term)
 taken at the points $x=v,\, y=u,\, \bar x=\bar u,\, \bar y=\bar v$
\beq\label{dphi}
D_\phi(x,y; \bar x,\bar y;z,\bar z)\,=\,-(K_{z,\bar z,x}\,+\,K_{z,\bar z,y})\,\delta(x-\bar x)\,\delta(y-\bar y)\,.
\eeq
We will also use
\begin{eqnarray}
&&J(x,y,u,v;k)\,\equiv\, \frac{\bar \alpha_s}{2\,\pi}\,
\int_{z,\bar z} \,e^{i\,k\,(z\,-\,\bar z)}\ J(x,y,u,v;z,\bar z)\\
&&J_\phi(x,y;k)\,\equiv\, \frac{\bar \alpha_s}{2\,\pi}\,
\int_{z,\bar z} \,e^{i\,k\,(z\,-\,\bar z)}\ J_\phi(x,y;z,\bar z) \\
&&D(x,y,u,v; \bar x,\bar y,\bar u, \bar v;k) \,\equiv\, \frac{\bar \alpha_s}{2\,\pi}\,
\int_{z,\bar z}\ \,e^{i\,k\,(z\,-\,\bar z)}\ D(x,y,u,v; \bar x,\bar y,\bar u,\bar v;z,\bar z) \\
&&D_\phi(x,y; \bar x,\bar y;k) \equiv \frac{\bar \alpha_s}{2\,\pi}\,
\int_{z,\bar z}\ \,e^{i\,k\,(z\,-\,\bar z)}\ D_\phi(x,y; \bar x,\bar y;z,\bar z)\,=\,
-\,\frac{\bar \alpha_s}{2\,\pi\,k^2}\ \delta(x-\bar x)\, \delta(y-\bar y)\,.
\end{eqnarray}

The last element we need to know in order to proceed is the "matrix element" ${\cal O}_0$ which characterizes the initial projectile wave function.

\subsubsection{The matrix element ${\cal O}_0$ }

In the dipole limit, the wave function of the incoming projectile is decomposed in the dipole basis and the $S$ - matrix $\Sigma^{PP}_0$ depends only on the dipole operator
\beq
\Sigma^{PP}_0[S]\,=\,\Sigma^{PP}_0[s_{x,y}]\,.
\eeq
This implies 
\beq
\Sigma^{PP}_0[S^\dagger\,\bar S]\,=\,\Sigma^{PP}_0[q^{s\bar s}_{x,y,y,x}]\,=\,\Sigma^{PP}_0[1\,+\,t_{x,y,y,x}]\,=\,\Sigma^{PP}_0[1\,+\,\phi_{x,y}]
\eeq
This expression can be expanded in Taylor series in $t$ ($\Sigma^{PP}_0[1]=1$):
\beq\label{texp}
\Sigma^{PP}_0[1\,+\,t_{x,y,y,x}]\,=\,
1\,+\,\sum_{m=1} \frac{1}{m!}\, \left\{\prod_{i=1}^m\,\int\,d^2x_i\,d^2y_i\,
t_{x_i,y_i,y_i,x_i}\,{\delta\over \delta t_{x_i,y_i,y_i,x_i}}\right\}\ \,\,
\Sigma^{PP}_0[1\,+\, t]|_{t=0}\,.
\eeq
For a projectile containing a single dipole at points $x,y$ the expression is very simple \beq
\Sigma^{PP}_0[1\,+\,t]\,=\,1\,+\,t_{x,y,y,x}\,=\,1\,+\,\phi_{x,y}\,.
\eeq

Given these ingredients, we can formulate the rules of calculation of multigluon production cross sections in terms of diagrams involving the propagation of the field $t$ in the external fields $s$ and $q$.

\subsection{Diagrammatics and Feynman rules}

In the dipole limit the generating functional $\cal Z$ reads
\bea\label{Zd}
{\cal Z}[j]\,=\,1&+&\int Ds\,Dq\,Dt\,\delta(t)\, W^T_{Y_0}[s,q]\ \times \nonumber \\
 &\times & \,e^{-\,\int_{Y_0}^{Y}\,d\eta\,\left[H_s\,+\,H_q\,+\,H_t-\,
\int_k {\cal O}_g(k)\,j\right]}\,
\left(\Sigma^{PP}_0[1\,+\,t]\,-\,2\,\Sigma^{PP}_0[s]\right)\,.
\eea
The $\Sigma^{PP}_0[s]$ term contributes only to the total cross section ($j=0$) which is not our interest in this paper.
Since this term does not depend on $t$ it vanishes under the action of $ {\cal O}_g$ and hence does not contribute
to particle production. We therefore omit $\Sigma^{PP}_0[s]$ in what follows. We also omit the unity which vanishes upon differentiation with respect to $j$.

It is convenient to get rid of $H_s$ and $H_q$ in the evolution operator by rewriting \eq{Zd} in the "interaction picture" with respect to the Hamiltonian $H_s+H_q$.
\beq\label{Zd1}
{\cal Z}[j]\,=\int Dt\, \delta(t) \,
e^{-\,\int_{0}^{Y}\,d\eta\,\left\{H_t[s_{\eta},q_{\eta}]\ -\
\int_k {\cal O}_g(k,\,[s_{\eta},q_{\eta}])\,j(k,\eta)\right\}}\,\Sigma^{PP}_0[1+t]\,.
\eeq
Here $s_\eta$ and $q_\eta$ are ``time'' dependent external fields, which are solutions of the corresponding equations of
motion with initial conditions $s_{\eta=0}=s$ and $q_{\eta=0}=q$. In the spirit of the mean field approximation we have omitted the factor $W^T_{Y=0}$ which  specifies the distribution of these initial conditions. This factor can be restored if desired.

It is now convenient to organize the calculation as perturbation theory in $t$. 
The $n$ - gluon production cross section is given by the path integral with $n$ insertion of the operator ${\cal O}_g(k,\,[s_\eta,q_\eta])$ and thus contains at most $n$ functional derivatives with respect to $t$. Since $t$ has to be set to zero at the end, this means that the rest of the expressions have to be kept only to order $t^n$. Thus we treat the homogeneous terms in the Hamiltonian $H_t$ \eq{Ht} as zeroth order. Their sum generates the "free" perturbative evolution, and this evolution has to be solved exactly. Just like with the evolution of $s$ and $q$ we see that any functional of $t$ in this framework evolves as
\beq
F_\eta[t]\,=\,F_{\eta=0}[t_\eta]
\eeq
where $t_\eta$ satisfies the equation
\begin{eqnarray}\label{dt}
\partial_\eta\, t_{x,y,u,v}&=&\frac{\bar \alpha_s}{2\,\pi}\,\int_{z} \left\{-\,[M_{x,y;z}\,+\,M_{u,v;z}\,-\,L_{x,u,v,y;z}]
\,\,t_{x,y,u,v}\,+\,L_{x,v,u,v;z}\,t_{x,y,u,z}\,s_{z,v}(Y-\eta)\,+
 \right.\,\nonumber \\
&+&
 L_{x,y,x,v;z}\,t_{z,y,u,v}\,s_{x,z}(Y-\eta)
\,+\,L_{x,y,u,y;z}\,t_{x,z,u,v}\,s_{z,y}(Y-\eta)\,+\,L_{u,y,u,v;z}\,t_{x,y,z,v}\,s_{u,z}(Y-\eta) \nonumber \\
&-&\left.
L_{x,y,u,v;z}\,(t_{z,y,u,z}\,s_{x,v}(Y-\eta)\,+\,t_{x,z,z,v}\,s_{u,y}(Y-\eta))
\right\}
\end{eqnarray}
 Note that since the evolution of $s$ and $t$ starts on the target side while the evolution of $t$ starts at the projectile side, the rapidity of the "background field" $s$ in \eq{dt} is $Y-\eta$. The same holds for the later discussion of the gluon emission operator. This has to be kept in mind although we will rarely indicate the rapidity label of $s$ and $q$ explicitly in the rest of this section.

\eq{dt} has to be solved with the initial condition
\beq
t_{\eta=0}( x,y,u,v)\,=\,t(x,y,u,v)\,.
\eeq
Formally the solution can be written as
\beq
t_{\eta}( \bar x,\bar y,\bar u,\bar v)\,=\,\int_{x,y,u,v}\,G_\eta(\bar x,\bar y,\bar u,\bar v|x,y,u,v)\ t(x,y,u,v)
\eeq
with the propagator
\beq
G_\eta(\bar x,\bar y,\bar u,\bar v|x,y,u,v)\,=\,\langle \bar x,\bar y,\bar u,\bar v |{\cal P} \exp\{-\int_0^\eta d\eta' h[s(Y-\eta'),q(Y-\eta')]\}|x,y,u,v\rangle
\eeq
where $h$ is the "first quantized" Hamiltonian that can be read off eq.(\ref{dt}). In other words the propagator $G$ is the solution of the differential equation
\begin{eqnarray}\label{Gt}
& &{\partial\over\partial\eta}\,G({\bf \bar x}|x,y,u,v)\,=\,\frac{\bar \alpha_s}{2\,\pi}\,
\int_{z} \left\{-\,[M_{x,y;z}\,+\,M_{u,v;z}\,-\,L_{x,u,v,y;z}]
\,G({\bf \bar x}|x,y,u,v|)\,+\, \right.\,\nonumber \\
&&+\,L_{x,v,u,v;z}\,G({\bf \bar x}|x,y,u,z)\,s_{z,v}\,+
 L_{x,y,x,v;z}\,G({\bf \bar x}|z,y,u,v)\,s_{x,z}
\,+\,L_{x,y,u,y;z}\,G({\bf \bar x}|x,z,u,v)\,s_{z,y}\,+\nonumber \\
&&+\,L_{u,y,u,v;z}\,G({\bf \bar x}|x,y,z,v)\,s_{u,z}
-\left.
L_{x,y,u,v;z}\,\left[G({\bf \bar x}|z,y,u,z)\,s_{x,v}\,+\,G({\bf \bar x}|x,z,z,v)\,s_{u,y}\right]
\right\}
\end{eqnarray}
with the initial condition at $\eta=0$
\beq
G_{\eta=0}({\bf \bar x}|x,y,u,v)\,=\,\delta_{x,\bar x}\,\delta_{y,\bar y}\,\delta_{u,\bar u}\,\delta_{v,\bar v}\,.
\eeq
Here ${\bf \bar x}\equiv(\bar x,\bar y,\bar u,\bar v)$.

We will also need the BFKL Pomeron Green`s function corresponding to propagation of the two point function $\phi$ (now ${\bf \bar x}=(\bar x,\bar y)$):
\beq\label{GBFKL}
{d\over d\eta}\,G^{BFKL}({\bf \bar x}|x,y)\,=\,\frac{\bar \alpha_s}{2\,\pi}\,
\int_{z} M_{x,y;z}\,\left[G^{BFKL}({\bf \bar x}|x,z)\,+\,G^{BFKL}({\bf \bar x}|z,y)\,-\,
G^{BFKL}({\bf \bar x}|x,y)\right]\,.
\eeq
The initial conditions for the BFKL propagator: at $\eta=0$
\beq\label{BFKLini}
G^{BFKL}_{\eta=0}({\bf \bar x}|x,y)\,=\,\delta_{x,\bar x}\,\delta_{y,\bar y}\,.
\eeq

The last term in eq.(\ref{Ht}) increases the power of $t$ and thus is to be treated as a perturbation. It defines the perturbation vertex $1\rightarrow 2$
\beq\label{lambda}
\int_{{\bf x},\ {\bf\bar x},\ {\bf\tilde x}}\ \lambda({\bf x},{\bf \bar x},{\bf\tilde x})\ t({\bf \bar x})\ t({\bf \tilde x}) \
{\delta\over \delta t({\bf x})}\,=\, -\, \int_{x,y,u,v,z}  \frac{\bar \alpha_s}{2\,\pi}\,L_{x,y,u,v;z}\ t_{z,y,u, z}\ t_{x, z,  z, v}
\ {\delta\over \delta t_{x,y,u,v}}\,.
\eeq

We note that when this operator acts on a functional of the two point function $\phi_{x,y}$ only, it produces another function of $\phi$ according to the rule (\eq{hamphi}) 
\beq
-\int_{ x,y,z} \frac{\bar \alpha_s}{2\,\pi}\,M_{x,y,z}\ \phi_{z,y}\ \phi_{x, z}
{\delta\over \delta \phi_{x,y}}
\eeq
which defines a triple $\phi$ vertex
\beq
\lambda^\phi(x,y,z)\ \equiv \ -\, \frac{\bar \alpha_s}{2\,\pi}\ M_{x,y,z}\,.
\eeq

Finally the gluon emission operator defines three types of insertions. The term $A_{-1}$ annihilates the propagating field $t$ with amplitude $J[s(Y-\eta),q(Y-\eta)]$, the term $A_0$ inserts the vertex $D[s(Y-\eta),q(Y-\eta)]$ on the propagator of $t$ and $A_1$ splits an existing propagator with the amplitude $V[s(Y-\eta),q(Y-\eta)]$.  

This defines the set of Feynman rules for the diagrammatic representation of the "partition function" $\cal Z$ eq.(\ref{Zd1}).
We denote each factor of $t$ in the expansion \eq{texp} by an empty circle. The propagator $G$ of \eq{Gt} is denoted by a straight line. The propagator carries an arrow indicating the direction of "time" Y (Fig. 1). 

\begin{figure}[htbp]
\centerline{\epsfig{file=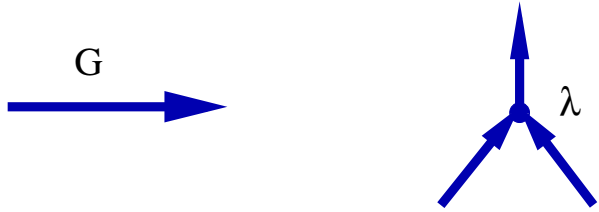,width=80mm}}
\caption{\it The propagator and the vertex $\lambda$}
\label{fig2}
\end{figure}
The vertex $\lambda$, \eq{lambda} is denoted by a full circle. Every vertex has two propagators entering and one propagator leaving (Fig.1)

Finally we denote insertions of $J$, $D$ and $V$ of \eq{ais},\eq{aiss},\eq{aisss} by a full circle with horizontal line. 
Note that the insertions of $J$ and $D$ are quite complicated, as they depend on rapidity dependent fields $s(Y-\eta)$ and $q(Y-\eta)$.
We cumulatively represent the factors $s$ and $q$ entering the insertion $J$ by dashed lines (see Appendix A for more details). 

To compute the $n$ gluons inclusive cross section we proceed as follows. 
\begin{enumerate}
\item At the projectile rapidity $Y$ draw $1\le m\le n$ open circles corresponding to the expansion of $\Sigma^{PP}_0$  \eq{texp} to 
$m$-th order (if the projectile is a single dipole there will be only one circle).

\item Each open circle is a sink for one propagator $G$ \eq{Gt}. Draw the diagrams of the field theory with propagator $G$ and 
vertex $\lambda$ of \eq{lambda} with $n$ insertions of the operators ${\cal O}_g(k_i)$ an rapidities $Y_0\,+\,Y_i$.

\item The only diagrams that contribute are the ones where each propagator starts on the source $J$ and ends either on the vertex $\lambda$ or on the vertex $V$ or on the open circle.

\item Finally we sum over all diagrams with $1\le m\le n$ and $N_V+N_J+N_D=n$, where $N_V$, $N_J$ and $N_D$ are the numbers of insertions of $V$, $J$ and $D$ respectively.

In each diagram the number of various vertices are related by 
\beq
m+N_{\lambda}+N_V=N_J
\eeq
where $N_\lambda$ is the number of vertices $\lambda$.

Since there are at most $n$ insertions of the source term $J$, the perturbation theory in $\lambda$ has to be developed at most to order $n-1$.

\end{enumerate}

In the next subsections we explicitly consider the examples of one-, two-, and three-gluon inclusive cross section.  For simplicity in all these examples we take the projectile as a single dipole $(x_0,y_0)$.

\subsection{Single inclusive gluon production}

\begin{figure}[htbp]
\centerline{\epsfig{file=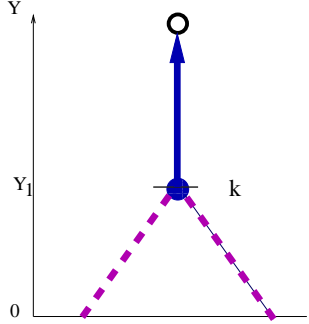,width=40mm}}
\caption{\it Single inclusive gluon production}
\label{fig3}
\end{figure}

Single gluon production rate is our first example.
Since $J=1$ and $m=1$ there is only one diagram (Fig. \ref{fig3}) which contributes to this observable.
Moreover the propagator entering this diagram is simply the BFKL propagator. The reason is that the projectile $\Sigma^{PP}_0$ is 
the functional only of the two point function $\phi_{x_0,y_0}$. On the other hand as discussed above, the two point function propagates independently of other components of $t$, \eq{hamphi}. The homogeneous part of \eq{hamphi} is simply the BFKL equation, and so the propagator of $\phi$ is the BFKL propagator.

The analytic expression for the diagram of Fig. 2 is
\begin{eqnarray}\label{Ogd}
{d\sigma\over dY_1\,dk^2}&=& \frac{\bar \alpha_s}{\pi}\,\int_{z,\bar z,x,y} \,e^{i\,k\,(z\,-\,\bar z)}\ G^{BFKL}(x_0,y_0|x,y;Y-Y_1)\ 
J(x,y;z,\bar z;Y_1)\nonumber\\
&=&\frac{\bar \alpha_s}{\pi}\,\int_{z,\bar z} \,e^{i\,k\,(z\,-\,\bar z)}\,\int_{x,y} {\cal N}_{x,z,y,\bar z}\,
G^{BFKL}(x_0,y_0|x,y;Y-Y_1)\times\nonumber \\
&\times&
\left[s^2_{z,\bar z}(Y_1)\,+\,s^2_{x,y}(Y_1)\,-\,
s^2_{z,y}(Y_1)\,-\,s^2_{x,\bar z}(Y_1)\right]\,.
\end{eqnarray}
This reproduces the result of Kovchegov and Tuchin \cite{Kovchegov:2001sc}.

\subsection{Double inclusive gluon production}

\begin{figure}[htbp]
\centerline{\epsfig{file=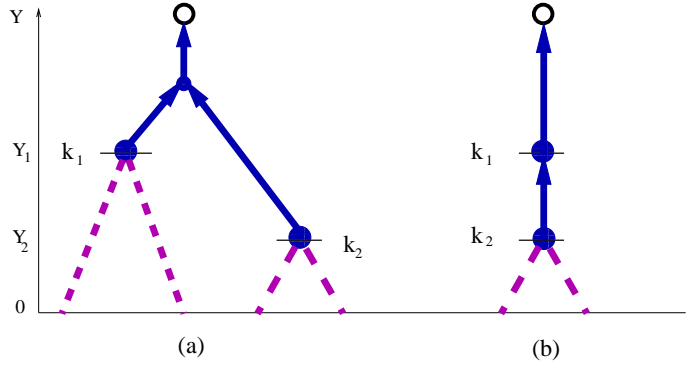,width=90mm}}
\caption{\it Double inclusive gluon production}
\label{fig5}
\end{figure}

For the double inclusive gluon production there are two diagrams depicted on (Fig. \ref{fig5}). The diagrams are topologically the same as those of \cite{braun}, however the propagators in Fig. 3 are not necessarily those of the BFKL Pomeron.
The diagramm on Fig.3a is given by 
\begin{eqnarray}\label{Ogdga}
{d\sigma\over dY_1\,dk_1^2 \,dY_2\,dk_2^2}|_{(a)}&=&\int_{0}^{Y-Y_1}d\eta\,
G^{BFKL}(x_0,y_0|x,y;\eta)\,\lambda^\phi(x,y,z)\,
\,G^{BFKL}(x,z|u_1,v_1;Y-Y_1-\eta)\nonumber \\
&\times& G^{BFKL}(z,y|u_2,v_2;Y-Y_2-\eta)\,\,
J_\phi(u_1,v_1;k_1;Y_1)\ J_\phi(u_2,v_2;k_2;Y_2)
\end{eqnarray}
where we used Einstein's summation convention for transverse coordinates $x,y,\, etc.$. 

The diagram on Fig. \ref{fig5}b is
\begin{eqnarray}\label{Ogdgc}
{d\sigma\over dY_1\,dk_1^2 \,dY_2\,dk_2^2}|_{(b)}&=&
G^{BFKL}(x_0,y_0|x,y;Y-Y_1)\, 
D(x,y,y,x|\bar x,\bar y,\bar u,\bar v; k_1;Y_1)\,\times
\nonumber \\
&& \times\, G(\bar x,\bar y,\bar u,\bar v|x_1,y_1,u_1,v_1;Y_1-Y_2)
\, J(x_1,y_1,u_1,v_1; k_2;Y_2)\,.
\end{eqnarray}

Note that in this diagram the BFKL propagator of two point function $\phi_{x,y}$ become a propagator of the three point function $\psi_{x,z,\bar z},\, \psi_{z,y,\bar z}\,$  $etc.$ below the insertion of the first gluon emission operator. This does not happen in the single gluon inclusive calculation nor in the diagram 
Fig. \ref{fig5}a. The propagation of the three point function couples to the "background fields" $s$ and $q$, and therefore this diagram violates the AGK cutting rules as stressed in \cite{JMK}.  There is one contribution to this diagram however which does not change the nature of $\phi$. This is the insertion of 
$D_\phi(x,y,\bar x,\bar y;k)$ of eq.(\ref{dphi}):
\bea\label{Ogdgb}
{d\sigma\over dY_1\,dk_1^2 \,dY_2\,dk_2^2}|_{(b;\phi)}&=&
G^{BFKL}(x_0,y_0|x,y;Y-Y_1)\ D_\phi(x,y;\bar x,\bar y;k_1) \ \times \nonumber \\ 
&\times& G^{BFKL}(\bar x,\bar y|u,v; Y_1-Y_2)\ J(u,v;k_2;Y_2)\,.
\eea
Nevertheless it does not seem useful to separate this particular contribution as it is not the only contribution even in the BFKL limit.

Our result coincides with that of \cite{JMK}. The detailed correspondence is as follows. 
The expression 
$$
\int_{0}^{Y-Y_1}d\eta\,
G^{BFKL}(x_0,y_0|x,y;\eta)\,\lambda^\phi(x,y,z)\,
\,G^{BFKL}(x,z|u_1,v_1;Y-Y_1-\eta)\ G^{BFKL}(z,y|u_2,v_2;Y-Y_2-\eta)
$$
is identical with $n_2$ of Ref. \cite{JMK}. Thus the contribution of diagram Fig.3a reproduces the first term in Eq. (32) of that reference.
\eq{Ogdgb} is identical with the very last term in Eq. (32) in \cite{JMK}, the term proportional to $M$
(without the factor $S$). 
The expression $$G^{BFKL}(x,y|u,v; Y_1-Y_2)\,J(u,v;k_2;Y_2)$$ is 
$M(x,x,y,Y_1;k_2,Y_2)+ (x\leftrightarrow y)$ of Ref. \cite{JMK}.
Finally \eq{Ogdgc} with \eq{Ogdgb} subtracted 
is the same as the ``$M\,S$'' terms of Eq. (32) in \cite{JMK}: the vertex $J(k_1,Y_1)$ brings in the
factor $s$ ($S$ of Ref. \cite{JMK}); the three-point function $M$ of \cite{JMK} equals
 $$G(\bar x,\bar y,\bar u,\bar v|x_1,y_1,u_1,v_1;Y_1-Y_2)\,
J(x_1,y_1,u_1,v_1; k_2;Y_2)$$
whereas the vertex $J(k_2,Y_2)$ equals $d$ in Eq. (27) 
of \cite{JMK}.
\subsection{Triple gluon inclusive production}
It is straightforward to draw Feynman diagrams which contribute to the triple gluon inclusive production cross section. There are nine distinct diagrams that are depicted in Fig. 4.  We do not write down analytic expressions for the different contributions: this is a straightforward if cumbersome endeavor. We only note that the new element that appears here relative to the single and double gluon production is the last diagram which involves insertion of the vertex $A_1$. Since the vertex $A_1$ turns a three point function $\psi$ into a generic four point function $t$, this diagram contains the general propagator $G$ below the rapidity $Y_1$. 
Thus even though the diagrams again are topologically  the same as in the BFKL Pomeron calculus \cite{braun}, the propagators are those of the generic four point function $t_{x,y,u,v}$ rather than BFKL propagators of $\phi_{x,y}$.

\begin{figure}[htbp]
{\epsfig{file=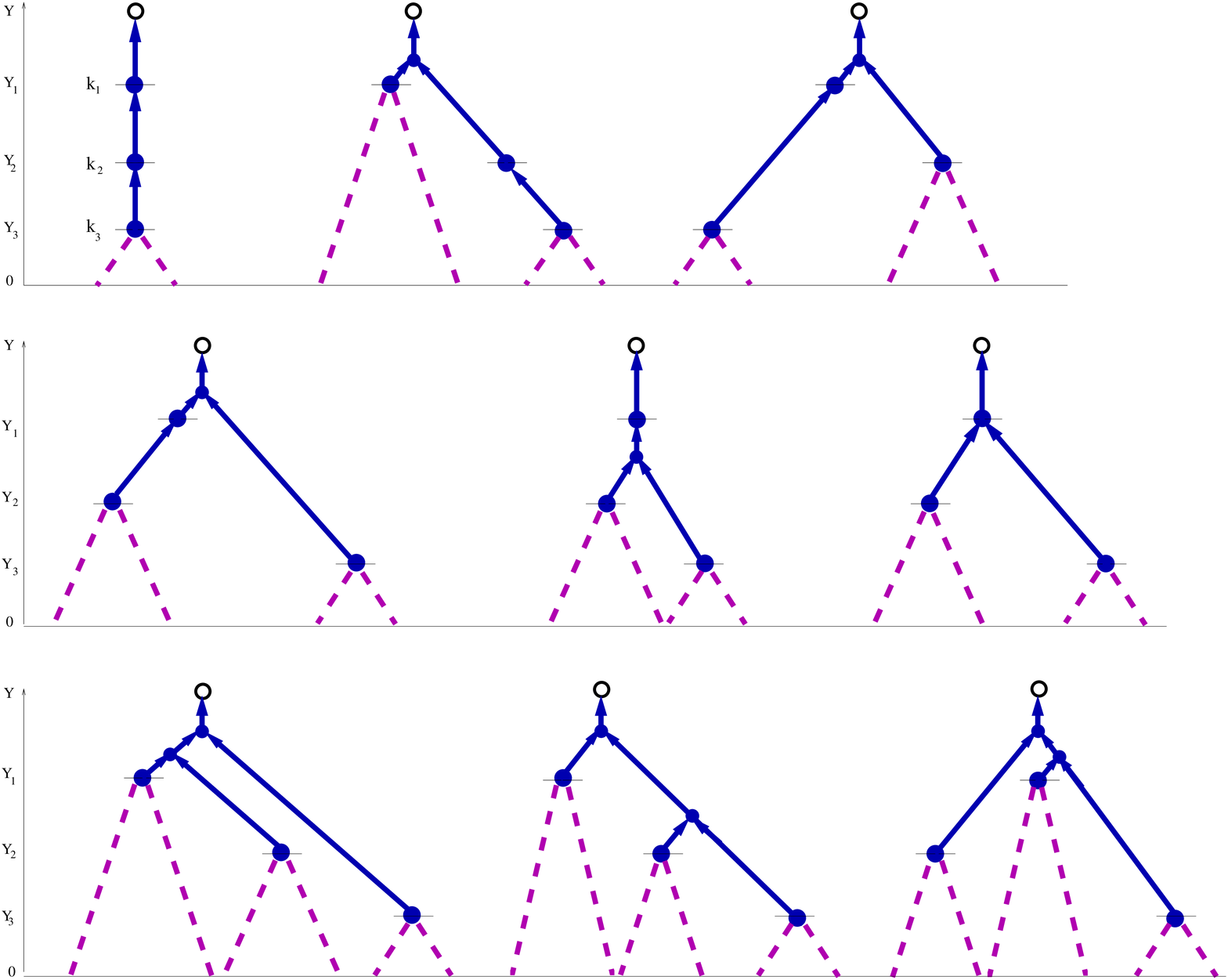,width=140mm}}
\caption{\it Three- gluon production}
\label{fig6}
\end{figure}

\subsection{Can we make sense of it?}
The diagrammatics involved in the calculation of multiple gluon production described in this section is simple, however an actual calculation would be quite involved. Single gluon production apart from the solution $s$ of the BK equation, requires only
the knowledge of the propagator of the two point function $\phi_{x,y}$ which is simply the BFKL propagator and thus is reasonably straightforward. The double inclusive gluon requires already the knowledge of the propagator of the three point function $\psi_{x,y,v}$. This we do not know in general since $\psi$ couples to the "background" field $s$. For three and higher gluon production we would need  the knowledge of the propagator of the generic four point function $t$ which is even more complicated. 
In this subsection we motivate a scheme which approximates the propagator of $t$ by a known function of the BFKL propagators. 
 
Let us first examine the evolution of the three-point function $\psi_{x,y,v}$ more closely. 
We rewrite the linear part of the equation of motion of $\psi$ here for convenience:
\begin{eqnarray}\label{psiagain}
\partial_\eta \,\psi_{x,y,v}&=&
\frac{\bar \alpha_s}{2\,\pi}\,\int_{z} \Big\{-\,{1\over 2}[M_{x,y;z}\,+\,M_{y,v;z}\,+\,M_{x,v;z}]
\,\,\psi_{x,y,v}\,
-\,L_{x,y,y,v;z}\phi_{z,y}\,s_{x,v}
\nonumber\\
&+&\,L_{x,v,y,v;z}\,\psi_{x,y,z}\,s_{z,v}\,+
 L_{x,y,x,v;z}\,\psi_{z,y,v}\,s_{x,z}-\,L_{x,y,y,v;z}\,\psi_{x, z, v}\Big\}\,.
\end{eqnarray}
We are interested in the high energy limit, where the target probed at any rapidity between $Y_0$ and $Y$ is black. The leading behavior of $\psi$ then is determined by taking the limit $s\rightarrow 0$.  
We have to be somewhat careful since 
$s(x,z=x)=1$ even in the deep saturation regime and this regulates the UV singularity in the virtual term  in the equation. Still to get a qualitative idea we can take $s=0$ but use $Q_s(Y-\eta)$ as the UV cutoff wherever a UV cutoff is required.
Taking naively $s=0$ in \eq{psiagain} we discover that only the first and the last terms survive. The first, virtual term is negative and UV divergent. Assuming that it gives the leading contribution to the evolution we find.
\beq
\psi_{x,y,v;\eta}\ \sim\ {\rm exp}\left\{\,-\,\frac{\bar \alpha_s}{2\,\pi}\eta\,\ln[ (x\,-\,v)^2\,Q_s^2(Y-\eta)]\ \right\}\,.
\eeq
This of course is only valid for $|x-v|>Q_s^{-1}(Y-\eta)$ since otherwise the last term in \eq{psiagain} gives a large contribution which cancels the putative UV divergence.
Note that the "source" term - the second term in \eq{psiagain} gives a competing contribution. At large distance $|x-v|>Q_s^{-1}(Y-\eta)$,  $s(x,v)$  
decays very steeply as \cite{LT}
\beq
s_{x,v}(Y-\eta)\propto {\rm exp}\left\{\,-\,\frac{\bar \alpha_s}{2\,\pi}(Y-\eta)\,\ln[ (x\,-\,v)^2\,Q_s^2(Y-\eta)]\right\}
\eeq
so that close to the projectile rapidity where $\eta<Y-\eta$ the contribution of this term is subleading. Closer to the target rapidity it dominates, still qualitatively the behavior is very similar. 
We thus conclude that while the two-point function $\phi$ grows with energy with the BFKL intercept, $\psi_{x,y,v}$ at large $|x-v|>Q_s^{-1}(Y-\eta)$ decreases with energy very quickly.

It is also possible understand the behavior of $\psi_{x,y,v}$ at small values of the separation $x-v$. For $x=v$, as explained above we have $\psi_{x,y,x}=\phi_{x,y}$. When the separation is smaller than $Q_s^{-1}$ but nonvanishing, we can estimate the behavior of $\psi$ by taking $s=1$ in \eq{psiagain}. This is obvious in the second term in \eq{psiagain}, since it is directly proportional to $s_{x,v}$. This also gives the leading contribution in the third (fourth) term, since the $z$- integral is peaked at $z=v$ ($z=x$). 
Setting $s=1$ in \eq{psiagain} we obtain the equation
\begin{eqnarray}
\partial_\eta\,\psi_{x,y,v}&=&
\frac{\bar \alpha_s}{2\,\pi}\,\int_{z} \Big\{-\,{1\over 2}[M_{x,y;z}\,+\,M_{y,v;z}\,+\,M_{x,v;z}]
\,\,\psi_{x,y,v}\,
-\,L_{x,y,y,v;z}\ \phi_{z,y}
\nonumber\\
&+&\,L_{x,v,y,v;z}\,\psi_{x,y,z}\,+
 L_{x,y,x,v;z}\,\psi_{z,y,v}\,-\,L_{x,y,y,v;z}\,\psi_{x, z, v}\Big\}
\end{eqnarray}
It is easy to check that this equation is solved by
\beq\label{psiphi}
\psi_{x,y,v}={1\over 2}(\phi_{x,y}+\phi_{y,v}-\phi_{x,v})
\eeq
if $\phi_{x,y}$ satisfies the linearized version of \eq{hamphi} - the BFKL equation.

In fact the relation \eq{psiphi} is required by the definition of $t_{x,y,u,v}$ \eq{shift} in the BFKL limit, where $S(x)=1+i\alpha(x)$ and $\bar S(x)=1+i\bar\alpha(x)$ are close to unity. Expanding $q^{s\bar s}_{x,y,y,v}$ to second order in $\alpha$ and $\bar\alpha$ and taking the correlator of $\alpha$ to be equal to correlator of $\bar\alpha$ (in accord with the discussion leading to \eq{simp}) we find the relation \eq{psiphi} directly from the definition of $t$. The same argument more generally gives
\beq\label{tphi}
t_{x,y,u,v}={1\over 2}\left(\phi_{x,u}+\phi_{y,v}-\phi_{x,v}-\phi_{y,u}\right)
\eeq
This again can be checked to solve the equation of motion \eq{dt} with $s=1$.

We stress that \eq{psiphi} and \eq{tphi} hold only in the linearized regime for $s=1$. Nevertheless this means that at small separations, $|x-v|<Q_s^{-1}$; $|y-u|<Q_s^{-1}$ the propagator of the four point function $t$ behaves as 
\begin{eqnarray}\label{propt}
&& G(x_0,y_0,u_0,v_0|x,y,u,v;\eta)=\\
&&{1\over 2}\left(G^{BFKL}(x_0,u_0|x,u;\eta)+G^{BFKL}(y_0,v_0|y,v;\eta)-G^{BFKL}(x_0,v_0|x,v;\eta)-G^{BFKL}(y_0,u_0|y,u;\eta)\right)\nonumber
\end{eqnarray}

Next let us take a closer look at the gluon emission vertex \eq{ais}, \eq{aiss}, \eq{aisss}. We will zoom into one of the terms in the vertex, but the same considerations exactly go through for all other terms. Consider one of the terms in $A_0$
\beq\label{a0}
({\cal N}_{y,z,v,\bar z}\,+\,{\cal N}_{x,z,u,\bar z}\,-\,{\cal N}_{x,z,v,\bar z}\,-\,{\cal N}_{y,z,u,\bar z})\,
q_{z,y,u,\bar z}\,t_{x, z,\bar z, v}{\delta\over \delta t_{x,y,u,v}}
\eeq 
Let us assume for a moment that the coordinates of the field $t_{x,y,u,v}$ entering this vertex are close together, so that $|x-v|<Q_s^{-1}$; $|y-u|<Q_s^{-1}$ where $Q_s$ is defined at the rapidity $Y-\eta$ of the quadrupole operator $q$ in \eq{a0}. Since $q_{z,y,u,\bar z}\ll 1$ for $|z-\bar z|>Q_s^{-1}$, the coordinates of the outgoing field $t$ are also pairwise within the same distance scale $Q_s^{-1}$. The same is true for all the other terms in $A_0$. 

Thus the vertex $A_0$ itself picks only the region of the phase space in the propagator $G$, in which the propagator can be taken as \eq{propt}.

The vertex $A_1$ is different in this respect, since the coordinates $z$ and $\bar z$ of the outgoing field are not necessarily close to each other. This vertex appears only for triple and higher gluon production.
For these multigluon observables it is therefore necessary to include the large distance suppression factor in the propagator itself. As we have discussed above this suppression is the property of the linearized evolution equation for $t$. In lieu of exact solution of this equation one can resort to modelling. The following form suggests itself as a reasonable approximation,:
\begin{eqnarray}\label{proptm}
&& G(x_0,y_0,u_0,v_0|x,y,u,v;\eta)={1\over 2}s_{x,v}(Y-\eta)s_{y,u}(Y-\eta)\times\\
&&\Big[G^{BFKL}(x_0,u_0|x,u;\eta)+G^{BFKL}(y_0,v_0|y,v;\eta) -G^{BFKL}(x_0,v_0|x,v;\eta)-G^{BFKL}(y_0,u_0|y,u;\eta)\Big]\nonumber
\end{eqnarray}
 It obviously has correct behavior both at small and large separations and the crossover in the behavior occurs on the correct distance scale.
 
To summarize, the diagrams for single gluon production contain only the propagator of $\phi$ and thus are the same as in the BFKL Pomeron calculus \cite{braun}. For double inclusive gluon production since the vertex $A_1$ does not appear, the leading contribution is again given by the diagrams where the propagators of $\psi$ can be taken at $s=1$\footnote{ There is a slight subtlety here, since the BFKL propagator leads to diffusion in transverse space. Thus one can worry about the fact that in Fig. 3b the transverse spread between the points in the propagator entering the vertex $J(k_2)$ is not necessarily small even if at the vertex $D(k_1)$ the spread was within $Q^{-1}_s$. If that where the case our argument about the reliability of the approximation \eq{propt1} would be invalid. However the diffusion in the BFKL equation is rather slow, so that the growth of radius is slower than the decrease of $Q_s(Y-\eta)$ as $\eta$ grows towards the target. It therefore follows that even allowing for diffusion, the transverse spread always stays small enough so that the approximation $s=1$ can be used for the propagator.}
\begin{eqnarray}\label{propt1}
&&G_\psi(x_0,y_0,v_0|x,y,v;\eta)=\\
&&{1\over 2}\left(G^{BFKL}(x_0,y_0|x,y;\eta)+G^{BFKL}(y_0,v_0|y,v;\eta)-G^{BFKL}(x_0,v_0|x,v;\eta)\right)\nonumber
\end{eqnarray}
In this approximation the propagators again become those of the BFKL Pomeron and the leading contribution preserves the AGK cutting rules, modulo the fact that the field $s$ couples directly in the vertex $D(k_1)$. This conclusion agrees with the argument of \cite{braun}. 

However for triple gluon production and higher we cannot use BFKL propagators to approximate the propagation of $\psi$ and $t$ in the external field $s$. Thus the interaction with $s$ appears not only at the gluon emission vertices and the AGK cutting rules are irretrievably lost. 

However, we believe that if the rapidities of all measured gluons are far from the target, a good quality approximation for all momenta $k$ except possibly $k_i\sim Q_s(Y-Y_i)$
can be obtained by using the Feynman rules with the vertices given in the previous subsections using as the propagator 
of the field $t$ \eq{proptm}.
The calculation then requires only the knowledge of the BFKL propagator, and naturally the solution of Kovchegov equation $s_{x,y}$.

\section{Discussion}
In this paper we have developed the formalism to calculate multigluon inclusive cross section at high energy for arbitrary number of gluons. The formulae in the full unabridged JIMWLK formulation are complicated and at the moment it is not clear how to develop a feasible calculational approach in this regime. We have also discussed the dipole model limit of the JIMWLK approach. In this case the expressions simplify dramatically. We have been able to develop a simple diagrammatic technique based on Feynman rules for calculations in this model. We have shown that for the double gluon inclusive production one recovers the results of \cite{JMK}. We note that the formalism can be extended easily to other observables. In Appendix B for example we consider gluon emission with fixed momentum transfer of the valence part of the projectile wave function.

We confirm that the AGK cutting rules are violated for multigluon production. The violation is due to the appearance of non Pomeron exchanges even within the dipole model. The calculation involves propagators of  three- and four point "multipoles" propagating in the external field $s$. 
However we presented an argument that for the double inclusive gluon production the leading high energy behavior is indeed given by the diagrams preserving the AGK cutting rules in accordance with conclusion of \cite{braun}.
On the other hand for three and higher gluons the AGK rules are genuinely violated.
 
Perhaps the most remarkable result of this paper is that only finite number of multipoles (namely multipoles which depend on two, three and four transverse coordinates) is needed for the calculation of any number of emitted gluons.

The propagators of these multipoles except for the dipole are not known analytically, but the problem is still on an entirely different level of complexity than the full JIMWLK. We have suggested an ansatz for the propagators based on its known UV and IR behavior which we believe should be a very good approximation for calculating the cross section.

It is interesting to note that although the vertex in the Hamitonian \eq{Ht} contributes only perturbatively to a fixed order when the number of gluons is fixed, one could in principle calculate other observables where the full evolution of $t$ will enter the fray. An example of such an observable could be average multiplicity. Since the evolution of $t$ is nonlinear and undoubtedly leads to saturation of growth of $t$ this suggests that the average multiplicity must be growing much slower with energy than in the pure BFKL calculation. This slowdown is of course affected by the saturation of $s$ but mostly by the saturation of $t$. It would be interesting to study this question in detail. In particular it is possible that classical solutions of the evolution equation for $t$ could be directly related to physical observables, much like classical solutions of field equations are related to particle production in QCD \cite{francois}.

\section*{Acknowledgments}

We are grateful to Heribert Weigert for discussions at the initial stages of this project.
The work of A.K. is supported by the Department of Energy under Grant No. DE-FG02-92ER40716. 

\appendix

\section{Appendix: Vertices  } \label{sec:A}
In this Appendix we collect explicit expressions for the vertices $J$ and $D$ which enter the gluon emission vertex eqs.(\ref{ais}, \ref{aiss}).
\begin{eqnarray}
J(x,y,u,v:z,\bar z)&=&({\cal N}_{y,z,v,\bar z} +\ {\cal N}_{x,z,u,\bar z})\,s_{x,y}\, s_{u,v}-({\cal N}_{x,z,v,\bar z}\,+\,{\cal N}_{y,z,u,\bar z})\,q_{x,y,u,v} 
\ \nonumber \\
&+&
({\cal N}_{x,z,v,\bar z}\,-\,{\cal N}_{y,z,v,\bar z})\,s_{x,z}\,q_{z,y,u,v}\,+\,
({\cal N}_{y,z,u,\bar z}\,-\,{\cal N}_{x,z,u,\bar z})\,s_{z,y}\,q_{x,z,u,v}\nonumber \\
&+&
({\cal N}_{x,z,v,\bar z}\,-\,{\cal N}_{x,z,u,\bar z})\,s_{\bar z,v}\,q_{x,y,u,\bar z}\,+
({\cal N}_{y,z,u,\bar z}\,-\,{\cal N}_{y,z,v,\bar z})\, s_{u,\bar z}\,q_{x,y,\bar z,v}
\nonumber \\
&+&({\cal N}_{y,z,v,\bar z}\,+\,{\cal N}_{x,z,u,\bar z}\,-\,{\cal N}_{x,z,v,\bar z}\,-\,{\cal N}_{y,z,u,\bar z})\,
q_{z,y,u,\bar z}\,q_{x, z,\bar z, v}
\end{eqnarray}
For the two point function $\phi$:
\begin{eqnarray}\label{jphia}
J_\phi(x,y;z,\bar z)&=&\,
{\cal N}_{y,z,x,\bar z}\,(s^2_{z,\bar z}\,+\,s^2_{x,y}\,-\,s^2_{x,z}\,-\, s^2_{y,\bar z})\,+{\cal N}_{x,z,y,\bar z}\,(s^2_{z,\bar z}\,+\,s^2_{x,y}\,-\,s^2_{z,y}\,-\,\,s^2_{\bar z,x})\,\\
&+&{\cal N}_{y,z,y,\bar z}\,(s^2_{y,\bar z}\,+\,s^2_{z,y}\,-\,s^2_{z,\bar z}\,-\,1)
+\,{\cal N}_{x,z,x,\bar z}\,(s^2_{\bar z,x}\,+\,s^2_{x,z}\,-\,s^2_{z,\bar z}\,-\,1)\,\nonumber 
\end{eqnarray}

\begin{figure}[htbp]
{\epsfig{file=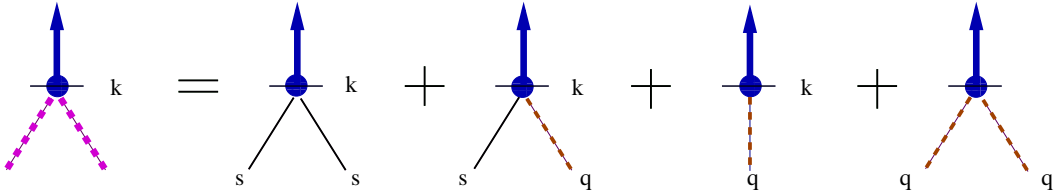,width=140mm}}
\caption{\it The "source" term  $J$ in the gluon emission vertex.}
\label{fig7}
\end{figure}

The vertex $D$ is given by
\begin{eqnarray}
&&D(x,y,u,v,\bar x,\bar y,\bar u,\bar v;z,\bar z)=({\cal N}_{x,z,v,\bar z}\,+\,{\cal N}_{y,z,u,\bar z})\delta_{x,\bar x}\delta_{y,\bar y}\delta_{u,\bar u}\delta_{v,\bar v}\\
&+&({\cal N}_{x,z,v,\bar z}\,-\,{\cal N}_{y,z,v,\bar z})\,s_{x,z}\,\delta_{\bar x,z}\,\delta_{\bar y,y}\,
\delta_{\bar u,u}\,\delta_{\bar v,v}\,
+\,
({\cal N}_{y,z,u,\bar z}\,-\,{\cal N}_{x,z,u,\bar z})\,s_{z,y}\,\delta_{\bar x,x}\,\delta_{\bar y,z}\,\delta_{\bar u,u}\,\delta_{\bar v,v}\,\nonumber\\
&+&
({\cal N}_{x,z,v,\bar z}\,-\,{\cal N}_{x,z,u,\bar z})\,s_{\bar z,v}\,
\delta_{\bar x,x}\,\delta_{\bar y,y}\,\delta_{\bar u,u}\,\delta_{\bar v,\bar z}\,
+
{\cal N}_{y,z,u,\bar z}\,-\,{\cal N}_{y,z,v,\bar z})\, s_{u,\bar z}\,
\delta_{\bar x,x}\,\delta_{\bar y,y}\,\delta_{\bar u,\bar z}\,\delta_{\bar v,v}\nonumber\\
&+&({\cal N}_{y,z,v,\bar z}\,+\,{\cal N}_{x,z,u,\bar z}\,-\,{\cal N}_{x,z,v,\bar z}\,-\,{\cal N}_{y,z,u,\bar z})(q_{z,y,u,\bar z}\,\delta_{\bar x,x}\,\delta_{\bar y,z}\,\delta_{\bar u,\bar z}\,\delta_{\bar v,v}
\,+\,q_{x, z,\bar z, v}
\,\delta_{\bar x,z}\,\delta_{\bar y,y}\,\delta_{\bar u,y}\,\delta_{\bar v,\bar z}
)\nonumber
\end{eqnarray}

Finally the vertex $V$ is given by 
\beq
V(x,y,u,v;z,\bar z)=({\cal N}_{y,z,v,\bar z}\,+\,{\cal N}_{x,z,u,\bar z}\,-\,{\cal N}_{x,z,v,\bar z}\,-\,{\cal N}_{y,z,u,\bar z})
\eeq
The vertex \eq{jphia} coincides with  $s$ in eq. (20) of  \cite{JMK}.\footnote{Note that the notations in this paper differ from those in \cite{JMK} so that the letter $s$ denotes a different quantity than in \cite{JMK}}

\section{Appendix: Multi-gluon production with momentum transfer} \label{sec:B}

It is straightforward to extend the formalism discussed in this paper to the case when in addition to emitted gluons, the transverse momentum transfer of the valence part of the projectile wave function is fixed. The only modification is that of the matrix element  ${\cal O}_0[S,\bar S] $. To impose momentum transfer $\bf p$ we take \cite{klw}
\beq\label{mt1}
{\cal O}_0[S,\bar S]_{\bf p}\,=\,\int d^2B\,d^2a\,e^{i\,{\bf p}\,a}\,
\Sigma^{PP}_0[S^\dagger(x)\,\bar S(x\,+\,a)]
\eeq
Here $B$ stands for impact parameter of the projectile.
In the dipole limit this becomes 
\beq\label{mt2}
{\cal O}_0[S,\bar S]_{\bf p}\,=\,\int d^2B\,d^2a\,e^{i\,{\bf p}\,a}\,
\Sigma^{PP}_0[1\,+\,t_{x,y,y+a,x+a}]
\eeq
One important difference from the formulae discussed above is that the four-point function $t$ enters already on the level of the single inclusive gluon cross section.

For a single dipole projectile ($B=(x_0+y_0)/2$)
the diagrams are topologically identical to those given in Section 4. The only difference is that the "sink" on which the propagators end is given by \eq{mt2}.
The expression for the single inclusive gluon production is 
\beq\label{mt3}
{dN\over dY_1\,dk^2\,\,d^2p}\,=\,\int d^2x_0\,d^2a\,e^{i\,{\bf p}\,a}\,G(x_0,y_0,y_0+a,x_0+a|x,y,u,v; Y-Y_1)
\,J(x,y,u,v;k,Y_1)
\eeq

We can also consider in this framework the case when the projectile consists of a single quark. The observable then corresponds to the process
$q\,+\,A \,\rightarrow\,q({\bf p})\,+\,X\,+\,g(k_1;Y_1)+\,g(k_2;Y_2)\,+\,...$.
To this process we take as the projectile single dipole  with the transverse coordinate of the antiquark at infinity:
\beq\label{mt4}
{\cal O}_0[S,\bar S]_{\bf p}\,=\,\int d^2x_0\,d^2a\,e^{i\,{\bf p}\,a}\,
[1\,+\,t_{x_0,y_0,y_0+a,x_0+a}]_{y_0\rightarrow \infty}
\eeq
The result for the single inclusive gluon production is
\beq\label{mt5}
{dN\over dY_1\,dk^2\,d^2p}\,=\,\int d^2x_0\,d^2a\,e^{i\,{\bf p}\,a}\,G(x_0,y_0,y_0+a,x_0+a|x,y,u,v)\,
\,J(x,y,u,v;k, Y_1)|_{y_0\rightarrow \infty}
\eeq
The propagator $G(x_0,y_0,y_0+a,x_0+a|x,y,u,v)|_{y_0\rightarrow \infty}$  diverges in the infrared and has to be
regularized. This is of course the reflection of the fact that the cross section for a color nonsinglet projectile is logarithmically IR divergent. If the gluon is emitted at the same rapidity as the valence quark the expression becomes
\beq\label{mt6}
{d\sigma\over dY_1\,dk^2\,d^2p}\,=\,\int d^2B\,d^2a\,e^{i\,{\bf p}\,a}
\,J(x_0,y_0,y_0+a,x_0+a;k,Y)|_{y_0\rightarrow \infty}
\eeq
which agrees with the result of \cite{BKNW}.

\end{document}